\documentclass[twocolumn]{aastex62}
\usepackage{natbib}
\usepackage{amsmath, mathtools}
\usepackage{CJK}

\received{}
\revised{}
\accepted{}
\submitjournal{ApJ}
\shorttitle{Galactic Center Astrometric Reference Frame}
\shortauthors{Sakai et al.}
\begin{document}

\begin{CJK*}{UTF8}{gbsn}

\title{The Galactic Center: An Improved Astrometric Reference Frame for Stellar Orbits around the Supermassive Black Hole}

\correspondingauthor{Shoko Sakai}
\email{shoko@astro.ucla.edu}

\author[0000-0001-5972-663X]{Shoko Sakai}
\affil{UCLA Department of Physics and Astronomy, Los Angeles, CA 90095-1547, USA}

\author[0000-0001-9611-0009]{Jessica R. Lu}
\affiliation{Astronomy Department, University of California, Berkeley, CA 94720, USA}

\author[0000-0003-3230-5055]{Andrea Ghez}
\affiliation{UCLA Department of Physics and Astronomy, Los Angeles, CA 90095-1547, USA}

\author[0000-0001-5341-0765]{Siyao Jia}
\affiliation{Astronomy Department, University of California, Berkeley, CA 94720, USA}

\author{Tuan Do}
\affiliation{UCLA Department of Physics and Astronomy, Los Angeles, CA 90095-1547, USA}

\author[0000-0003-2618-797X]{Gunther Witzel}
\affiliation{UCLA Department of Physics and Astronomy, Los Angeles, CA 90095-1547, USA}

\author[0000-0002-2836-117X]{Abhimat K. Gautam}
\affiliation{UCLA Department of Physics and Astronomy, Los Angeles, CA 90095-1547, USA}

\author[0000-0002-2186-644X]{Aurelien Hees}
\affil{SYRTE, Observatoire de Paris, Universit\'e PSL, CNRS, Sorbonne Universit\'e, LNE, 61 avenue de l’Observatoire 75014 Paris}
\affil{UCLA Department of Physics and Astronomy, Los Angeles, CA 90095-1547, USA}

%\author[0000-0002-6753-2066]{M.R. Morris}
%\affil{UCLA Department of Physics and Astronomy, Los Angeles, CA 90095-1547, USA}

\author{E. Becklin}
\affil{UCLA Department of Physics and Astronomy, Los Angeles, CA 90095-1547, USA}

\author{K. Matthews}
\affil{Astrophysics, California Institute of Technology, MC 249-17, Pasadena, CA 91125, USA}

\author[0000-0003-2874-1196]{M. W. Hosek Jr.}
\affil{Institute for Astronomy, University of Hawaii, Honolulu, HI 96822, USA}

\begin{abstract}

Precision measurements of the stars in short-period orbits around the supermassive black hole at the Galactic Center are now being used to constrain general relativistic effects, such as the gravitational redshift and periapse precession.  One of the largest systematic uncertainties in the measured orbits has been errors in the astrometric reference frame, which is derived from seven infrared-bright stars associated with SiO masers that have extremely accurate radio positions, measured in the Sgr A*-rest frame.
We have improved the astrometric reference frame within $14''$ of the Galactic Center by a factor of 2.5 in position and a factor of 5 in proper motion.
In the new reference frame, Sgr A* is localized to within a position of 0.645 mas and proper motion of 0.03 mas yr$^{-1}$.
We have removed a substantial rotation (2.25 degrees per decade), that was present in the previous less-accurate reference frame used to measure stellar orbits in the field. With our improved methods and continued monitoring of the masers, we predict that orbital precession predicted by General Relativity will become detectable in the next $\sim$5 years.

\end{abstract}

\keywords{adaptive optics, astrometry, black hole, Galactic Center}

\section{Introduction} \label{sec:intro}

Over the past 23 years, infrared astrometric and radial velocity data have been gathered for stars orbiting the supermasssive black hole (SMBH) at the center of the Galaxy \citep[e.g.][]{Ghez:2008ty,Gillessen:2009fn}.  The diffraction-limited speckle and adaptive optics observations obtained from the W.~M. Keck Observatory and the Very Large Telescope (VLT) have enabled us to characterize the SMBH associated with Sgr A* with unprecedented accuracy.  The mass of the SMBH has been estimated to be $4.02 \pm 0.16 \times 10^6 M_{\odot}$  at a distance of $7.86 \pm 0.14$ kpc \citep{Boehle:2016ko}, which is based on the orbits of two stars, S0-2 and S0-38. 
In 2018, the star S0-2, with an orbital period of $\sim$16 years, has gone through closest approach, enabling the first measurement of the gravitational redshift at the Galactic Center  (Do et al (in preparation), \cite{2018GRAVITY}).
Furthermore, in $\sim$ 5 years, 
the detection of the periapse precession predicted by General Relativity will be detectable if a reference frame stability of $\sim0.02$ mas yr$^{-1}$ can be achieved \citep{Weinberg2005}. Thus, the need for an improved astrometric reference frame is especially timely.

To make precise and accurate measurements of the stellar orbits around the SMBH, the sky-plane positions of stars around the SMBH are monitored using near-diffraction-limited observations from 8--10 m telescopes over many years. There are approximately 3,000 stars detected within a 10\arcsec \; radius of the Galactic Center.
However, since nearly all stars are moving within the observed field of view, it is challenging to transform the observed relative astrometry from each observation into a coordinate system that ties multi-epoch observations together.

The astrometric reference frame for the Galactic Center stellar orbits that was first used was based on the cluster-rest frame method, in which a set of reference stars were assumed to have no net motion.  However, the method was limited by the intrinsic dispersion of the cluster itself which did not improve with time \citep{Yelda:2010ig}. 
A better method for establishing an astrometric reference frame in the IR was suggested by \citet{Menten1995} and \citet{Reid2003} and later adapted by \citet{Yelda:2010ig,Ghez:2008ty,Gillessen:2009fn}.  We utilize radio-emitting SiO maser stars whose positions and proper motions have been determined precisely with respect to the radio source associated with the SMBH, Sgr A*. While Sgr A* is bright at radio wavelengths, it is faint and easily confused in the IR.  In contrast, the SiO masers are bright in both the IR and radio and can be utilized as near-perfect astrometric calibrators to bridge the radio and IR astrometric solutions.  
Because none of the masers are found in the central $10''$ region around the Galactic center (the field of view used for the dynamical study of the SMBH, hereafter referred to as the {\em central 10'' field}), it is necessary to mosaic the surrounding region of $\sim 22''$, such that the IR astrometric data of seven masers can be extracted.  By matching the IR positions of the  masers to their corresponding radio positions, the IR astrometric reference frame of the Galactic Center region is established. The presumed dynamical center, the massive black hole at Sgr A*, should then be at the origin of the reference frame and at rest.  

The reference frame in this Galactic Center region was previously determined in 2008, based on only three epochs of SiO maser data \citep{Ghez:2008ty}, and later improved in 2010 with six epochs of observations \citep{Yelda:2010ig}. Additional three epochs of maser data (2011 - 14) were included in the reference frame applied in \citet{Boehle:2016ko}.
The IR reference frame is constructed with the assumption that Sgr A* is at rest at the origin, deteremined from the radio astrometric observations.   
%the accuracy of the reference frame is quantified by mean of the 
%Sgr A*'s offset from zero in both position and proper motion in the IR. 
The precision of the reference frame is then characterized by how well the IR positions of the masers agree with the corresponding radio positions.
The objective is to make the agreement as close to zero as possible, and in particular, to assure that the reference frame does not display any shift with time, which is present in the orbit of S0-2 \citep[e.g.][]{Boehle:2016ko}.
Based on the analysis of the astrometric data obtained with NIRC2 at Keck Observatory, \citet{Yelda:2010ig} reported that the reference frame was determined with an accuracy of $0.81$ mas in position and $0.17 $ mas yr$^{-1}$ in proper motion. Based on the observations of eight masers, \citet{Plewa2015} reported an accuracy of $\sim 0.17$ mas and $\sim 0.07$ mas yr$^{-1}$ in positions and proper motion respectively.
%\textcolor{red}{We should probably put their uncertainties here from the two papers as we will refer to them later.}. 

In this paper, we present a new astrometric IR reference frame.  Since \citet{Yelda:2010ig}, seven additional epochs of maser mosaic data have been obtained.  In addition, several modifications were made in various stages of the reference frame construction process including the use of  updated maser proper motions at radio wavelengths, the use of an improved source extraction tool, AIROPA (Anisoplanatic and Instrumental Reconstruction of Off-axis PSFs for AO), to extract astrometric positions on maser mosaic frames, and an improved method of mosaicking lists of stars detected in each epoch of observation.  In Section~\ref{sec:data}, we present the data used to derive the absolute reference frame and how the starlists were created.  The method of constructing the astrometric reference frame is detailed in Section~\ref{sec:ref_frame} and the results are presented in \ref{sec:Results}.  The dependence of the reference frame on various methods utilized and modified since \citep{Yelda:2010ig} are discussed in Section~\ref{sec:discussion}.  The paper by Jia et al (2018, in preparation) and this paper both focus on the application of the astrometry of NIRC2 data for the Galactic Center orbit study.  This paper concentrates on investigating the stability of the astrometric reference frame reflected by the position and the proper motion of Sgr A* in IR, while Jia et al (2018) examines how to refine the procedure of cross-epoch alignment of astrometric data.

\section{Data} \label{sec:data}

%\subsection{Radio Observations}
%{\bf Do we need a separate subsection when we did not do the radio obs?}

\subsection{Radio Observations}

Maser stars that are detected both in the IR and radio are used to bring the IR stellar positions into an Sgr A* (radio) rest frame.  \citet{Reid2007} presented radio data of masers which were based on five epochs of observations over the period of 1995 -- 2006.  In this paper, we use radio positions and proper motions from M. Reid (2018, private communication) that have been updated with two additional epochs of observations.   
%\textcolor{red}{Don't we need to say something about the difference in the reduction of the radio with the rotation removed?}
%\todo[size=\small]{Present the radio measurements in a table.}

\subsection{Infrared Observations}

\begin{figure*}[htb!]
\epsscale{0.7}
\plotone{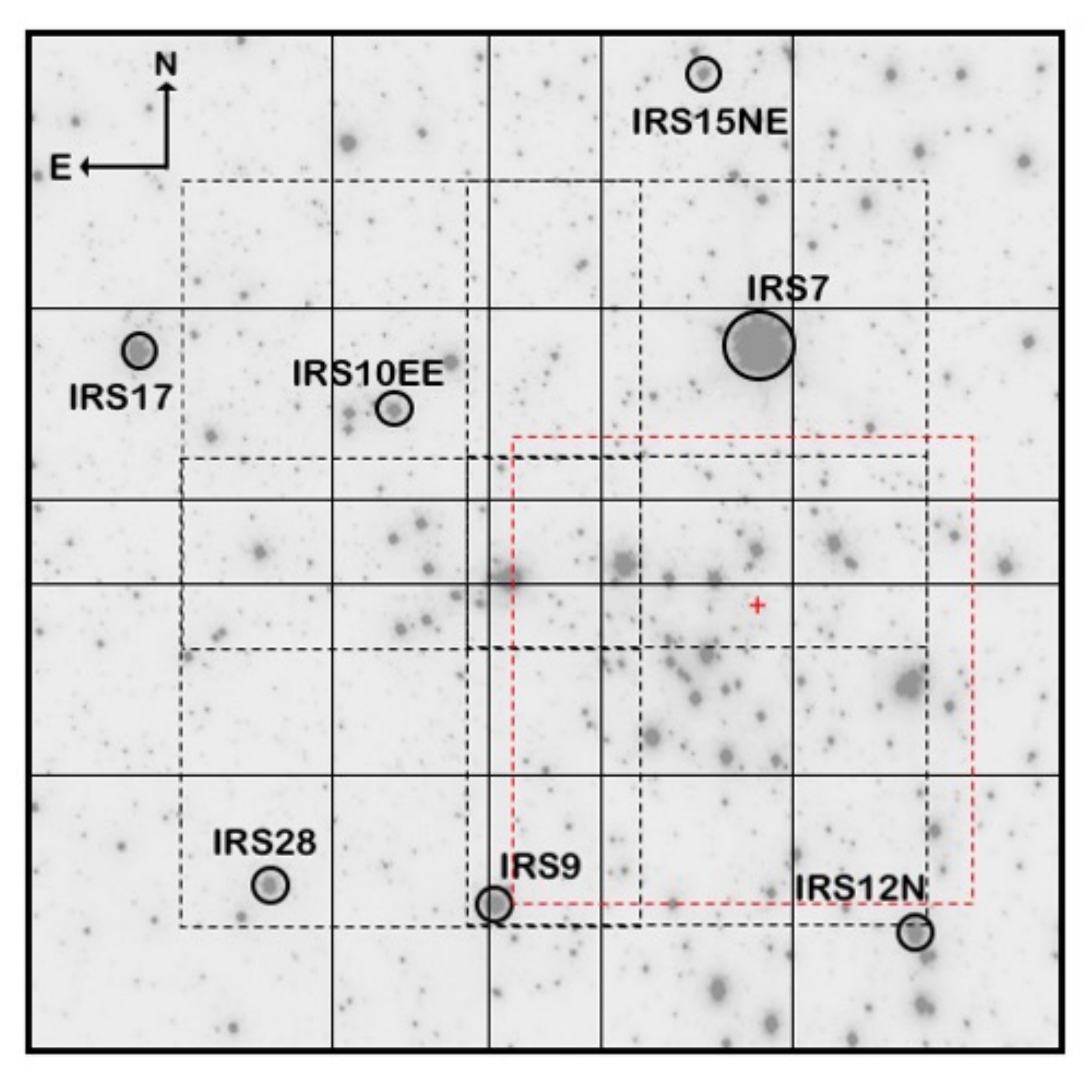}
\caption{NIRC2 K$^{\prime}$ image of the Galactic Center showing the maser positions in solid black circles.  The 3$\times$3 dither positions of the mosaic are shown by solid-line squares, and the 2$\times$2 dither positions by dashed lines. The position of Sgr~A* is indicated by a red cross and the central-10 image used for the orbit study is shown by dashed red box.  Each NIRC2 field is $10\arcsec \times 10\arcsec$.  The entire maser mosaic field is $22\arcsec \times 22\arcsec$. 
%$\textcolor{red}{FROM JLU: I am not sure if that red cross is actually on the position of Sgr A*. Also, could we add a box of a different color to mark the central 10" field of view?} $ 
% on the laptop:  GalacticCenter/RefFrame/Papers/fig_map.pptx
\label{figure:map}}
\end{figure*}

All the data for the IR observations of SiO masers in the Galactic Center region were obtained using the laser guide-star adaptive optics \citep[LGSAO;][]{vanDam2006,Wizinowich2006} facility on the Keck II telescope at the W. M. Keck Observatory. Images were obtained with a near-infrared facility imager (NIRC2; PI: K. Matthews) through the K$^\prime$ 
($\lambda_0 = 2.12\;\micron, \Delta \lambda = 0.35 \;\micron$) bandpass.   The camera has 1024 $\times$ 1024 pixels, with a plate scale of 9.952 mas per pixel (\citet{Yelda:2010ig}).  Since 2014, the AO system and NIRC2 camera were realigned, and the plate scale was changed to 9.971 mas per pixel (\citet{Service2016}).  All maser observations were performed with the position angle set to zero.
USNO 0600-28577051, which is offset by $9.4"$E and $69.5"$N from Sgr A*, served as the tip-tilt star for all of the LGSAO observations.
%\textcolor{red}{UPDATE WITH POST-2014 PLATE scale from Max Service's paper.}. 
%This is the same set-up as the one that is always used for the central 10$\arcsec$ observation that is used to monitor stellar orbits around the Sgr A* \citep{Ghez:2008ty,Boehle:2016ko}.
%The masers, however, are located outside this central 10$\arcsec$ area.
The integration time for each exposure on the Galactic Center masers was 10.86 s for most epochs, each comprised of 60 co-added 0.181 s exposures in order to avoid saturating the bright masers. 
%For the 2010 observing run, the exposure time was set to 1.81 s, each comprised of 10 co-added 0.181 second exposures. 
The integration time for the central 10$\arcsec$ field is much longer than the maser observations, with each exposure being 28 s, because this field was chosen to both include Sgr A* and to avoid bright stars that would saturate, such that a longer exposure time was possible. 

We used a widely-dithered ($6\arcsec \times 6\arcsec$) 9-point box pattern with multiple images at each of the nine positions.  The number of exposures per dither position changed throughout the 13 years of observations, varying from 3 to 18.   Starting in 2012, additional observations of a 4-point box pattern (with a smaller dither of $3\arcsec \times 3\arcsec$), with multiple images per dither position, were also obtained in order to average down astrometric errors from unaccounted for field variability in the point-spread function (PSF) due to atmospheric and instrumental aberrations. The data from the 4-point box pattern had not been incorporated in any previous analysis.  
%\textcolor{red}{[JLU: What about the data with different PAs? Aren't we including it now?]}
Figure~\ref{figure:map} shows the two mosaic patterns covering an area large enough to include seven masers closest to the GC.  Figure~\ref{figure:map} also shows the positions of the seven masers.   The full field of the maser mosaic is $22\arcsec \times 22\arcsec$, corresponding to $\sim$0.88 pc $\times$ 0.88 pc at the distance of $R_0 = 8.0$ kpc.  
%Although \citet{Reid2007} presented radio observations of 15 masers, only seven of them were detected on our mosaicked NIRC2 field.
A summary of the IR observations and the delivered image quality is given in Table~\ref{table:obs}. 
\citet{Yelda:2010ig} included the maser observations from 2005 -- 2010.  \citet{Boehle:2016ko} then incorporated three additional epochs of maser observations spanning 2011--2013.  However, they did not include the inner 2 $\times$ 2 dither positions in the reference frame construction.   
The new maser observations added in this paper since \citet{Boehle:2016ko} are indicated by asterisks in Table~\ref{table:obs}.
%, which includes the four-point positions taken in 2013 and 2014, and 13 positions taken annually from 2014 to 2017.  %Compared to \citet{Yelda:2010ig}, which was the last in-depth analysis of the GC reference frame from our group, seven additional epochs of data have been incorporated in the study presented in this paper.

\begin{deluxetable*}{lccccccc}
%\tablecolumns{9}
\tabletypesize{\scriptsize}
\tablewidth{0pt}
\tablecaption{Summary of GC Maser Mosaic Images \label{table:obs}}
\tablehead{
  \colhead{Date} & 
  \colhead{$t_{exp,i}$ x Co-Add} & 
  \colhead{$N_{exp}$\tablenotemark{a}} & 
  \colhead{$K^{\prime}_{\mbox{lim}}$\tablenotemark{b}} & 
  \colhead{FWHM\tablenotemark{c]}} & 
  \colhead{Strehl\tablenotemark{d}} & 
  \colhead{$N_{stars}$\tablenotemark{e}} & 
  \colhead{$\sigma_{pos}$\tablenotemark{f}} \\ 
  \colhead{(UT)} & 
  \colhead{(s)} & 
  \colhead{per dither position} & 
  \colhead{(mag)} & 
  \colhead{(mas)} & 
  \colhead{} & 
  \colhead{} & 
  \colhead{(mas)} \\
}
\startdata
2005 Jun 30 & 0.181 x 60 & 2 (3x3) & 15.18 & 60.1 & 0.32 & 1143 & 1.25 \\
2006 May 3  & 0.181 x 60 & 3 (3x3) & 15.57 & 61.2 & 0.28 & 1242 & 1.10 \\
2007 Aug 12 & 0.181 x 60 & 3 (3x3) & 15.66 & 56.6 & 0.29 & 1474 & 1.15 \\
2008 May 15 & 0.181 x 60 & 3 (3x3) & 15.92 & 53.5 & 0.35 & 1861 & 1.05 \\
2009 Jun 28 & 0.181 x 60 & 9 (3x3) & 16.12 & 63.5 & 0.26 & 2274 & 1.10 \\
2010 May 4 & 0.181 x 10 & 18 (3x3) & 15.52 & 67.6 & 0.25 & 1110 & 1.25 \\
2011 July 20 & 0.181 x 60 & 6 (3x3) & 15.87 & 63.2 & 0.27 & 2033 & 1.20 \\
2012 May 15 & 0.181 x 60 & 6 (3x3) + 3 (2x2)$\tablenotemark{*}$& 15.82 & 55.9 & 0.34 & 2226 & 1.10 \\
2013 July 2 & 0.181 x 60 & 18 (3x3) + 4 (2x2)$\tablenotemark{*}$ & 16.11 & 59.1 & 0.31 & 2745 & 1.20 \\
2014 May 20 & 0.181 x 60 & 18 (3x3)$\tablenotemark{*}$ + 3 (2x2)$\tablenotemark{*}$ & 16.14 & 65.8 & 0.24 & 2586 & 1.20 \\
2015 July 10 & 0.181 x 60 & 18 (3x3)$\tablenotemark{*}$ + 3 (2x2)$\tablenotemark{*}$ & 16.44 & 65.4 & 0.27 & 3216 & 1.20 \\
2016 May 15 & 0.181 x 60 & 18 (3x3)$\tablenotemark{*}$ + 3 (2x2)$\tablenotemark{*}$ & 15.99 & 80.1 & 0.19 & 2329 & 1.30 \\
2017 May 5 & 0.181 x 60 & 18 (3x3)$\tablenotemark{*}$ + 18 (2x2)$\tablenotemark{*}$ & 16.51 & 58.9 & 0.31 & 3265 & 1.20 \\
\enddata
\tablenotetext{a}{Number of exposures per dither position. The dither pattern is indicated in brackets.}
\tablenotetext{b}{$K^{\prime}_{\mbox{lim}}$ is the magnitude at which the cumulative distribution function of the observed $K^{\prime}$ magnitudes reaches 90\% of the total sample size.}
\tablenotetext{c}{Mean FWHM of stars detected in each epoch.}
\tablenotetext{d}{Mean Strehl ratio of stars detected in each epoch.}
\tablenotetext{e}{Number of stars detected in each epoch.}
\tablenotetext{f}{Mean total astrometric measurement error.}
\tablenotetext{*}{Maser data which had not been used in previous GCOI analyses.}

\tablecaption{K'$_{lim}$ needs a figure caption.}
\end{deluxetable*}

\subsection{IR Data Reduction}

The new NIRC2 data were reduced following the methods described in \citet{Yelda:2014by}.
Each NIRC2 frame was sky-subtracted, flat-fielded, corrected for bad pixels and also corrected for  the effects of optical distortion and differential atmospheric refraction \citep{Yelda:2010ig,Service2016}.  The sky background was estimated by taking the median of sky exposures nightly.    
For optical distortion correction, two independent lookup tables are used.  The AO was realigned in 2015, thus for the data taken prior to 2015, the optical distortion correction determined by \cite{Yelda:2010ig} is applied, while those taken in 2015 and later, the correction by \cite{Service2016} is used.

%\textcolor{red}{Need pipeline reference. Need distortion solution reference (\citet{Service2016})}.  
For each dither position, a bright isolated star is selected whose position is used to register and mosaic together multiple exposures.  The final image for each dither position was created by including only those frames having a FWHM less than 1.25 times the smallest FWHM value measured for the dither position.  If the number of exposures per dither position was only 3, which was the case for most of the earlier observations, and for the 2$\times$2 dither pattern for all years except 2017, all exposures were used to create the final image at each dither position.  Furthermore, the frames for each dither position were subdivided into three independent subsets.  Each subset consisted of frames of similar FWHM and Strehl ratios.  The images in each subset were then combined to make three ``submap'' images for each dither position.  The standard deviation of measured positions over the three submap images was used for initial estimates of astrometric uncertainties.  The positional uncertainties estimated from submaps range from 0.06 pixels ($\sim$0.6 mas) on the average for stars brighter than K$^{\prime}$=12 up to $\sim 0.1$ pixel for stars $14<K^{\prime}<15$.   The signal-to-noise ratio is $\sim$10,000 for $K \sim 12$ stars.  For fainter $K \sim 14-15$ mag stars, the S/N ratio is $\sim$ 4,000.
The astrometric uncertainties are larger, in general, in the maser mosaic data than in the Central 10\arcsec data.   This is because the Central 10 observations is much longer and deeper; the Central 10 data are comprised of hundreds of frames, each with the total exposure time of 2.8 sec $\times$ 10 coadds.  

%\textcolor{red}{[JLU: This last sentence seems like it should come after the Starlists section. Also, we need to say PA=0 and define X and Y as $\sim$R.A. and Dec. Finally, should we say something about the astrometric errors being larger, in general, in our maser mosaic data than in the central 10'' data and why?]}

\subsection{Starlists \label{sec:starlists}}
With the fully reduced dithered images, the next step is to make a master starlist consisting of values of stellar positions and proper motions.  Instead of creating a mosaicked image, we first extract stellar positions from the combined images at each dither position and then mosaic the starlists as described below \citep{Yelda:2014by}.

First, the photometry and astrometry were extracted on each dither position for each epoch using the AIROPA package developed by the UCLA Galactic Center group (\citet{Witzel2016}), based on the PSF-fitting code StarFinder (Diolaiti et al. 2000). 
%resubmission
This package was designed to take atmospheric turbulence profiles, instrumental aberration maps, and images as inputs and then fit field-variable PSFs to deliver improved photometry and astrometry on crowded fields. 
Although a single PSF was applied for the data used in this paper, the PSF extraction benefited from a much improved method as AIROPA uses improved StarFinder subroutines.
For each dither position, we selected $\sim 30-60$ isolated stars spread throughout the NIRC2 field of view, which are combined to create a single PSF model. This PSF is then used to extract positions and fluxes using PSF-fitting methods. A similar analysis is performed on the three submap images as well.

The starlists from all the dither positions were then combined together to create one master starlist for each epoch, by following the iterative steps described below. 
(1) First, we use the positions of stars presented in Table~6 of \cite{Boehle:2016ko} (hereafter GC starlist) as a starting point to launch an iterative process of mosaicking together starlists from individual mosaic positions to create a combined single master starlist for each epoch of NIRC2 observation.  The GC starlist is comprised of $\sim$830 stars, spanning the same region as the maser frames.
The positions given in arcseconds and proper motion values given in mas yr$^{-1}$ had been determined from the previous construction of the reference frame, in which accelerating sources were excluded (see \S\ref{sec:ref_frame} for details).  
%The stars had been flagged as accelerating if their $\chi^2$ values in proper motion fits to their x and y positions as a function of time were larger than a certain cutoff value.  This will be explained in more detail in Section 3.   
%xxx Explain how you knew which are those accelerating sources, maybe refer to a table of them XXX
An initial guess for the positions of stars in any given epoch can then be estimated using the values given in this GC starlist. 
% JLU: Redundant?: Even though the goal is to update the proper motion values of the GC starlist, the positions estimated using the values in the previous GC starlist serve as good starting point. 
Matching and transforming a starlist from one dither position to the GC starlist is  performed by fitting a third-order, 20-parameter (10 in each dimension) polynomial to the positional offsets of $\sim$ 100 stars.
%resubmission
The third-order fit was used in order to take care of the time-dependent residuals, likely stemming from the variable PSF, that were present after the geometric distortion corrections (\cite{Yelda:2010ig} \& \cite{Service2016}) were applied.  More details are given in Appendix~\ref{app:ldc}. 
This transformation step was iterated as each iteration added more stars, and thus a better transformation was calculated.
%using a tool called FlyStar, which is described briefly in REF.  A third-order, 20-parameter (10 in each dimension) polynomial is fit to the positional offsets of $\sim$100 stars on each dither position in order to determine the best transformation. 
%\textcolor{red}{[JLU: Didn't we iterate here??]} 
This matching procedure is repeated for all observed dither positions. 

(2) Now we have either 9 or 13 transformed starlists in arcsec for each epoch, depending on how many dither positions were observed (see Table~\ref{table:obs}). 
Next, all the starlists for a given epoch are combined together to create one master-list, spanning $\sim 22 \times 22$ arcsec per epoch.  For stars found in the overlap regions, the mean positions were used, with the standard deviation of the mean used as the positional uncertainties.  %resubmission
The typical astrometric uncertainties in the mosaicked image range from $\sim 0.05$ mas for stars brighter than $K \sim 12$ mag, up to $\sim 2$mas for $K \sim 15$ mag.  For comparison, the mean uncertainties in the Central 10$^{"}$ data, which are much deeper than the maser mosaic observations, are around $\sim$ 0.09 mas.

(3) The newly created mosaicked master-list is used as the reference starlist and steps 1 and 2 above is repeated six times.  After three or four iterations, the standard deviation of the match between the reference starlist and the mosaicked starlist for each mosaic position converges to $<$1 mas ($\sim$0.1 pixel).  

The accuracy of the stellar positions after the first iteration in the Step 2 is 1.7 mas and 2.2 mas in the East and North respectively, which decrease to 1.3 and 1.8 mas for subsequent iterations.
The uncertainty in the N/S direction is worse than in the E/W. This is likely due to the position of the tip tilt star located NNE of the entire maser mosaic field, which causes the PSF to be elongated in that direction and leads to poorer measurement precision.   
%Starlists from each dither position are matched to this masterlist by using a second-order polynomial fit.  maybe give the exact formula for this polynomial explicitly XXX Steps 2 and 3 are repeated 11 times, and the resulting starlist is used as the master starlist for the epoch.  
The numbers of stars and mean position uncertainties for each epoch are listed in Table~\ref{table:obs}.

\begin{deluxetable*}{lrccccccc}
\tabletypesize{\scriptsize}
\tablewidth{0pt}
\tablecaption{Astrometry of SiO Masers}
\tablehead{
  \colhead{} & 
  \colhead{} & 
  \colhead{} & 
  \colhead{$T_0$} & 
  \colhead{$T_0$\tablenotemark{b}} & 
  \colhead{RA Position\tablenotemark{c}} & 
  \colhead{DEC Position\tablenotemark{c}} & 
  \colhead{RA Velocity\tablenotemark{d}} & 
  \colhead{DEC Velocity\tablenotemark{d}} \\ 
  \colhead{Maser} & 
  \colhead{$K'$} & 
  \colhead{$\tilde{\chi}^{2}$\tablenotemark{a}} & 
  \colhead{IR} & 
  \colhead{IR+Radio} & 
  \colhead{[IR - Radio]} & 
  \colhead{[IR - Radio]} & 
  \colhead{[IR - Radio]} & 
  \colhead{[IR - Radio]} \\ 
  \colhead{} & 
  \colhead{(mag)} & 
  \colhead{} & 
  \colhead{(year)} & 
  \colhead{(year)} & 
  \colhead{(mas)} & 
  \colhead{(mas)} & 
  \colhead{(mas yr$^{-1}$)} & 
  \colhead{(mas yr$^{-1}$)} \\
}
\startdata
           IRS 9  &  9.063  & 0.17  & 2011.1  & 2010.2  &  0.139 $\pm$  1.007 $\pm$  0.219  &  0.313 $\pm$  1.019 $\pm$  0.375  &  0.073 $\pm$ 0.027 $\pm$ 0.043  &  0.088 $\pm$ 0.042 $\pm$ 0.074 \\
           IRS 7  &  7.658  & 0.52  & 2011.0  & 2009.4  &  1.557 $\pm$  1.044 $\pm$  5.001  &  2.767 $\pm$  1.044 $\pm$  5.002  &  0.034 $\pm$ 0.076 $\pm$ 0.122  & -0.829 $\pm$ 0.067 $\pm$ 0.225 \\
         IRS 12N  &  9.538  & 0.03  & 2011.1  & 2006.6  & -0.253 $\pm$  1.001 $\pm$  0.210  & -0.199 $\pm$  1.003 $\pm$  0.409  & -0.022 $\pm$ 0.009 $\pm$ 0.027  &  0.000 $\pm$ 0.017 $\pm$ 0.052 \\
          IRS 28  &  9.328  & 0.12  & 2011.6  & 2011.2  &  0.328 $\pm$  1.006 $\pm$  0.588  &  0.231 $\pm$  1.016 $\pm$  0.428  & -0.080 $\pm$ 0.024 $\pm$ 0.128  & -0.089 $\pm$ 0.036 $\pm$ 0.099 \\
        IRS 10EE  & 11.270  & 0.11  & 2011.1  & 2008.8  &  0.106 $\pm$  1.006 $\pm$  0.173  & -0.651 $\pm$  1.012 $\pm$  0.196  &  0.026 $\pm$ 0.024 $\pm$ 0.024  & -0.002 $\pm$ 0.037 $\pm$ 0.028 \\
        IRS 15NE  & 10.198  & 0.03  & 2011.0  & 2005.0  &  0.118 $\pm$  1.002 $\pm$  0.320  &  0.268 $\pm$  1.003 $\pm$  0.307  & -0.009 $\pm$ 0.014 $\pm$ 0.040  &  0.021 $\pm$ 0.016 $\pm$ 0.038 \\
          IRS 17  &  8.910  & 0.24  & 2011.1  & 2007.9  & -1.457 $\pm$  1.005 $\pm$  1.855  &  0.331 $\pm$  1.023 $\pm$  1.167  & -0.356 $\pm$ 0.012 $\pm$ 0.603  &  0.268 $\pm$ 0.038 $\pm$ 0.214 \\
%           IRS 9  &  9.06  & 0.12  & 2010.6  & 2010.1  &  0.016 $\pm$  0.100 $\pm$  0.217  &  0.390 $\pm$  0.165 $\pm$  0.373  &  0.058 $\pm$ 0.023 $\pm$ 0.043  &  0.041 $\pm$ 0.041 $\pm$ 0.074 \\
%           IRS 7  &  7.66  & 0.77  & 2011.0  & 2009.5  &  1.312 $\pm$  0.421 $\pm$  5.000  &  2.858 $\pm$  0.323 $\pm$  5.002  &  0.061 $\pm$ 0.105 $\pm$ 0.122  & -0.914 $\pm$ 0.074 $\pm$ 0.225 \\
%         IRS 12N  &  9.54  & 0.04  & 2011.3  & 2006.8  & -0.212 $\pm$  0.069 $\pm$  0.211  & -0.262 $\pm$  0.097 $\pm$  0.411  &  0.002 $\pm$ 0.012 $\pm$ 0.027  &  0.024 $\pm$ 0.019 $\pm$ 0.052 \\
%          IRS 28  &  9.33  & 0.10  & 2011.9  & 2011.2  &  0.407 $\pm$  0.121 $\pm$  0.580  &  0.254 $\pm$  0.136 $\pm$  0.422  & -0.084 $\pm$ 0.023 $\pm$ 0.128  & -0.035 $\pm$ 0.027 $\pm$ 0.099 \\
%        IRS 10EE  & 11.27  & 0.18  & 2011.0  & 2009.0  &  0.151 $\pm$  0.143 $\pm$  0.174  & -0.790 $\pm$  0.158 $\pm$  0.197  &  0.063 $\pm$ 0.030 $\pm$ 0.024  &  0.004 $\pm$ 0.036 $\pm$ 0.028 \\
%        IRS 15NE  & 10.20  & 0.06  & 2011.3  & 2005.4  &  0.106 $\pm$  0.084 $\pm$  0.323  &  0.319 $\pm$  0.092 $\pm$  0.309  & -0.025 $\pm$ 0.017 $\pm$ 0.040  &  0.011 $\pm$ 0.019 $\pm$ 0.038 \\
%          IRS 17  &  8.91  & 0.26  & 2011.2  & 2007.9  & -1.493 $\pm$  0.146 $\pm$  1.868  &  0.500 $\pm$  0.258 $\pm$  1.170  & -0.290 $\pm$ 0.030 $\pm$ 0.603  &  0.246 $\pm$ 0.062 $\pm$ 0.214 \\
\tableline
Weighted Average\tablenotemark{e}  &         & 0.17  &         & 2008.5  &  0.015 $\pm$  0.458  &  0.033 $\pm$  0.455  &  0.005 $\pm$ 0.018  &  0.004 $\pm$ 0.025 \\
%\tablenotemark{e}  &         & 0.22  &         & 2008.6  &  0.010 $\pm$ 0.040 $\pm$ 0.106 & -0.156 $\pm$ 0.138 $\pm$ 0.051 &  0.018 $\pm$ 0.002 $\pm$ 0.008 &  0.004 $\pm$ 0.004 $\pm$ 0.011 \\
\enddata
\tablecomments{Infrared (first) and radio (second) formal uncertainties are reported for each maser's position and velocity. Average distortion errors ($\sigma\sim$1 mas) for each maser are added in quadrature to the infrared formal uncertainties. X and Y increase to the east and north, respectively.}
\tablenotetext{a}{$\tilde{\chi}^2$ is the average of the X and Y $\chi^2$ per degree of freedom.}
\tablenotetext{b}{Average $T_0$ from both IR and radio measurements weighted by velocity errors.}
\tablenotetext{c}{Positional offsets computed for the common epoch of 2008.6. The first and second uncertainties are those in the IR and radio respectively.}
\tablenotetext{d}{Velocity offsets computed for the common epoch of 2008.6. The first and second uncertainties are those in the IR and radio respectively.}
\tablenotetext{e}{Weighted average and error in the weighted average are reported for all columns except the $\tilde{\chi}^2$ and $T_0$ columns, where we report the average.}
\label{tab:maser_polyfit}
\end{deluxetable*}

\section{Sgr A*-Radio Rest Reference Frame} \label{sec:ref_frame}

\begin{figure*}[ht!]
%\epsscale{0.8}
\plotone{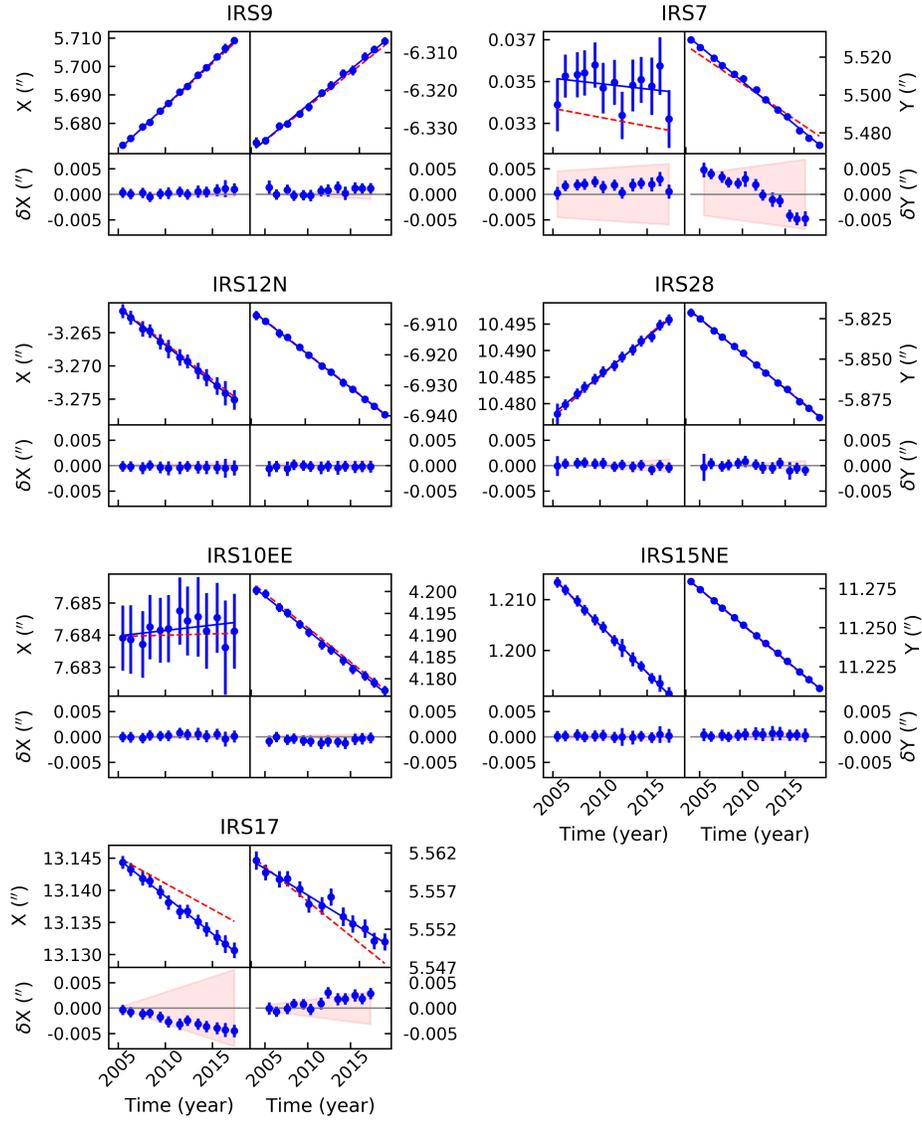}
\caption{The positions over time for the 7 masers, one per panel, in both the X/East ({\em top left}) and Y/North ({\em top right}) directions.  The IR and radio proper motion fits are over-plotted as {\em blue} and {\em red} lines, respectively.  The 1$\sigma$ error limits in the radio proper motions are shown by the red shaded regions. 
In the bottom panels, the residuals of the IR observed points with respect to the radio velocities are plotted for each maser.  The error bars show the IR positional uncertainties while the red shaded region represent the uncertainty in the radio velocities.
\label{fig:xpos}
}
%/u/shoko/code/python/refframe/xposALL.py
\end{figure*}

\subsection{Constructing the IR Astrometric Reference Frame}
The goal of constructing the IR astrometric reference frame is to produce a set of secondary astrometric standard stars whose proper motions are linear and well measured.  Their positions at a given epoch can then be used to define the coordinate system for that epoch.  Stellar positions from multiple epochs can then be aligned together in the IR reference frame  to derive the proper motions and accelerations of those stars in the region close to Sgr A* for orbits determinations.  

We start out with 13 starlists, each produced from one epoch of NIRC2 mosaic observations of SiO masers by following the procedure described in the previous Section. The 13 sets of data span 13 years from 2005 to 2017.  First, the seven maser positions from the individual starlist per epoch need to be transformed to their radio positions propagated to the corresponding epoch.  The radio maser positions are calculated from the proper motion measurements given by M. Reid (private comm).

%resubmission
We note that the intrinsic size of Sgr A* is less than 1 AU, corresponding to $\sim$ 0.13 mas at the distance of the Galactic Center \citep{ReidBrunthaler2004}.  Thus the uncertainty in the reference frame resulting from the error in the size measurement of Sgr A* is minimal.  In addition, the position of Sgr A* in the radio is determined with an uncertainty of $\sim$1 mas.
We also note that Sgr A*'s proper motion is $-3.151 \pm 0.018$ mas/yr and $5.547 \pm 0.026$ mas/yr in the east and south directions respectively \citep{ReidBrunthaler2004}, with respect to background QSO.  However, the reference frame being derived in this paper is based on the assumption that Sgr A* is at rest. In other words, our frame is relative to Sgr A*.

Compared to the radio data used in \citet{Yelda:2010ig}, one or two additional astrometric points for each maser were included in the derivation of proper motions. For the radio positions and proper motions, it is assumed that the Sgr A* is at the center of the Galactic Center and at rest. Thus ideally the same assumption holds for the IR reference frame.  The positional uncertainties in the propagated radio measurements vary from 0.3 mas to 5 mas depending on the maser, with a mean of $\sim 1.35$ mas.   For the alignment of IR and radio masers positions, six independent parameters were used (translation, rotation and pixel scale independently in the East and North, to first order).  The transformed IR starlist is now in the astrometric reference frame in which the Sgr A* radio source is at rest.  This exercise was repeated for all 13 epochs, resulting in a set of 13 starlists that are all in the same radio astrometric reference frame.  

The errors in the transformation of the IR positions to the radio Sgr A* rest frame in each epoch were calculated by a jackknife method, in which one maser was dropped from the alignment at a time.   The standard deviation of seven ``drop-one-maser'' cases for each star was adopted as the positional uncertainty.  The total astrometric uncertainty for a star's position in every epoch is estimated by combining the positional, alignment and distortion correction errors in quadrature.  For the distortion correction error, the average value of 1 mas, as derived by \cite{Yelda:2010ig} and \cite{Service2016} was added to the East and North astrometric errors.  

Next, the proper motions for all astrometric secondary standard stars are determined.  The 13 epochs of starlists are in a common coordinate system and are cross-matched to identify each star across all epochs. The proper motions are then determined by fitting a linear velocity model to each stars' positions over time. 
%In the fit, the distortion correction uncertainty of 1 mas was added to the the East and North astrometric errors, in quadrature, of all measurements in all epochs. 
The proper motion errors were estimated using a jackknife resampling method.   Out of $\sim$3900 stars detected in the entire maser mosaic field, 1008 stars were detected in 12 epochs or more, out of 13 epochs.  For a star to be included in the secondary astrometric standard star list, it needed to meet the following conditions:  the magnitude of proper motion both in the x and y direction $<$10 mas/yr, the uncertainties in the x and y proper motions $<$0.2 mas/yr, and both $\chi^2$ values in x and y directions for proper motion fits are less than 20, to exclude accelerating stars. The final secondary standard star list is comprised of 748 stars.  Table~\ref{tab:secondary} lists the properties of secondary IR astrometric standard stars.

\section{Results} \label{sec:Results}

The quality and stability of the astrometric reference frame is quantified in a number of ways.
First, Table~\ref{tab:maser_polyfit} shows the summary of how well the IR and radio SiO maser positions and proper motions agree.  The columns of the table are:  (1) name of the maser used; (2) $K^{\prime}$ magnitude; (3) total reduced $\chi^2$ values for proper motion fits; (4) the average time of the IR positional measurements; (5) the epoch at which the positional difference is expected to have the smallest uncertainty for each maser, estimated using Equation~3 in \citep{Yelda:2010ig}; (6) \& (7) the mean difference in the East and North direction between the measured IR and predicted radio positions. The IR formal uncertainties are dominated by the average distortion error of 1 mas which is added in quadrature; and (8) \& (9) differences between the IR and radio proper motions.  In the last row, the weighted average of these residuals are tabulated.  For a perfect reference frame, these residuals should be zero both in position and velocity; thus they are useful for estimating the stability of the reference frame and how well the positions and proper motions of Sgr A* are determined in the IR reference frame. 

The comparisons listed in Table~\ref{tab:maser_polyfit} show that the position of Sgr A* in the IR is known to within 0.458 mas and 0.455 mas in the East and North direction, respectively, in the year $2008.5$.  The velocity of Sgr A* in the IR reference frame is estimated with an accuracy of $\sim 0.03$ mas/yr.  The results of how well the IR maser positions agree with the radio positions are also shown in Figure~\ref{fig:xpos}.  For each maser, the observed IR positions are plotted as a function of time in the top panels, with the fitted IR velocities over-plotted.  
The coordinates are defined such that X increases to the East and Y increases to the North.
Also shown by red dashed lines are radio velocities provided by M. Reid (private comm).  In the bottom panels, the residuals of the IR observed points with respect to the radio velocities are plotted for each maser.  The error bars show the IR positional uncertainties while the red shaded region represent the uncertainty in the radio velocities.  %The IR positional errors are dominated by the uncertainty in distortion correction which is 0.001 arcsec.   
The mean weighted RMS of the IR fits to the radio velocities is 0.35 mas yr$^{-1}$. 

We note that when comparing the positions of the maser sources in the IR and radio, the intrinsic source positions may differ.  SiO maser emission originates in the extended atmospheres of late red giants and supergiants.  
The emission comes from a radius of $\sim$4AU typically, corresponding to $\sim 0.5$mas at the distance of the Galactic Center \citep{Reid2007}; and the SiO emission may not be symmetrically distributed. The resulting systematic difference between the maser emission centroid and the photospheric centroid should be significantly less than this. 
The radio centroid of the emission typically has a measurement uncertainty of about $\pm 0.5$ mas \citep{Reid2007}, which is larger than the expected systematic uncertainty.  
Even for the largest maser in our sample, IRS~7, its photospheric diameter is estimated to be 2.6 mas \citep{Pott_IRS7}, with a ratio of molecular envelope radius to the photospheric size or $\sim$2 - 2.2
\citep{Perrin_AGB,Ohnaka_AGB,Danchi_AGB,Wittkowski_AGB}. We expect very little systematic offset in the final position of Sgr~A*, since seven masers are used in establishing the reference frame, and thus any uncertainties arising from the radio-IR mismatch of their centroids should be random.  
It is very unlikely that in all seven masers, the SiO emission ring is skewed or asymmetric in the same direction.

\begin{figure}[ht!]
\plotone{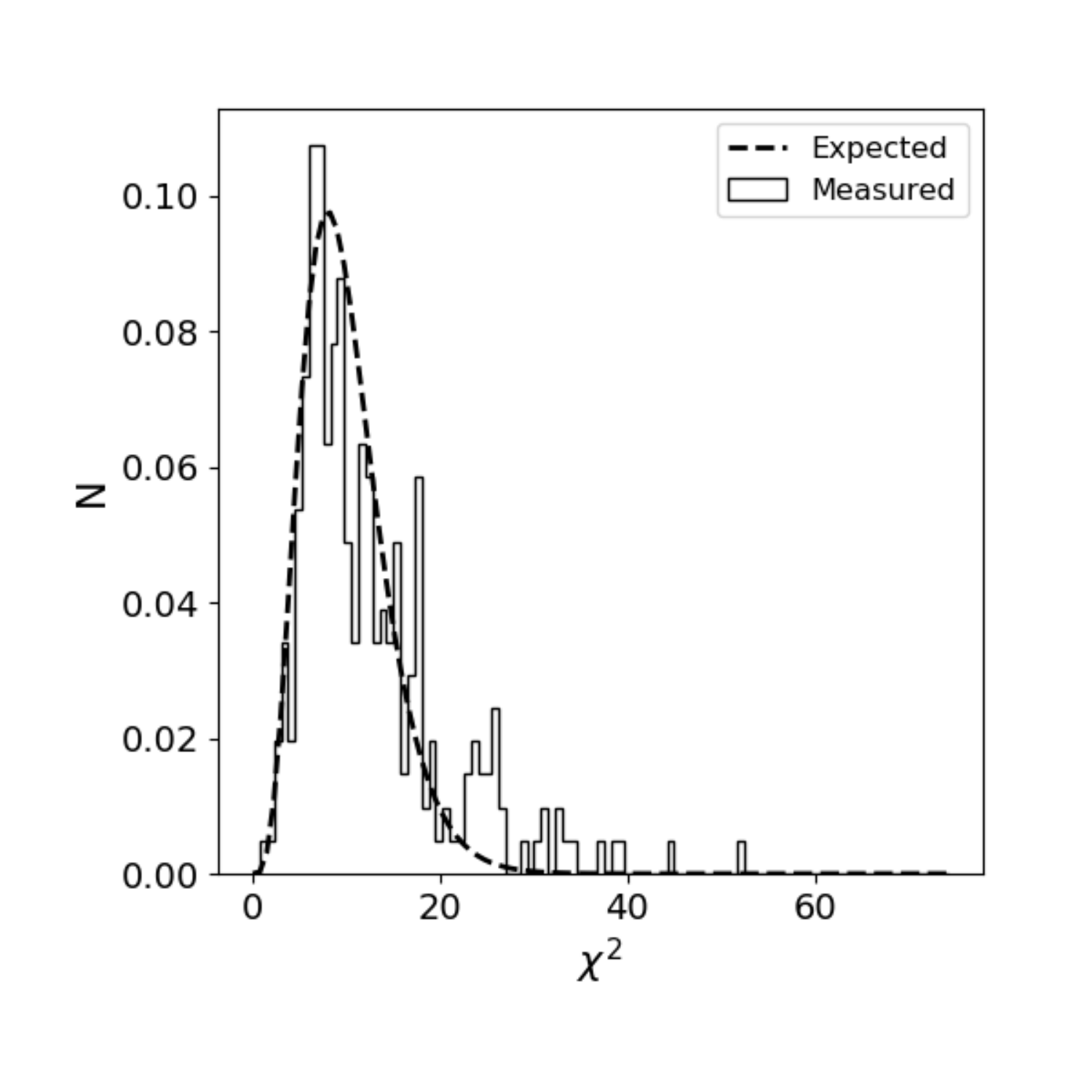} 
\caption{
Distribution of $\chi^2$ values of velocity fits to all the stars found on the maser mosaic field, before applying cutoffs to create the list of secondary astrometric starndard stars. The expected distribution is that of a corresponding $\chi^2$ distribution for the degree of freedom of 10, for stars found in at least 12 out of the total of 13 epochs.  For a star to be selected as a secondary astrometric standard star, its $\chi^2$ value in both $V_{RA}$ and $V_{DEC}$ must be less than 20.
\label{fig:chi2} }
\end{figure}

Because the velocity-fit $\chi^2$ values are used to select the astrometric standard stars, we need to be certain that the errors are estimated correctly.  If they are underestimated, then more stars would be excluded based on their inaccurately-estimated large $\chi^2$ values, even though these stars are not likely accelerating.  Overestimated errors would, on the other hand, lead to truly accelerating sources being included in the astrometric standard starlist, as their $\chi^2$ values are underestimated.  Examining the $\chi^2$ distribution as shown in Figure~\ref{fig:chi2}, the astrometric errors seem to be estimated accurately as they follow the $\chi^2$ distribution of the corresponding degree of freedom which is shown by the solid line.  

% TABLE ASTROMETRIC AND PSF STARS 
% on laptop  /Users/shokosakai/GalacticCenter/code/refframe/Papers
% PSF data in subdir PSF_table/

\begin{deluxetable*}{lrrrrrrrrrrccc}[htb!]
\tabletypesize{\scriptsize}
\tablewidth{0pt}
\tablecaption{Galactic Center Secondary IR Astrometric Standard and PSF Stars \label{tab:secondary}}
\tablehead{
  \colhead{} &
  \colhead{} &
  \colhead{} &
  \colhead{} &
  \colhead{} &
  \colhead{} &
  \colhead{} &
  \colhead{} &
  \colhead{} &
  \colhead{} &
  \colhead{}  &
  \colhead{}   &
  \colhead{Astrometric} &
  \colhead{} \\
    \colhead{Name} &
  \colhead{K'} &
  \colhead{$T_{0,IR}$} &
  \colhead{Radius} &
  \colhead{$\Delta$ R.A.} &
  \colhead{$\sigma_{R.A.}$\tablenotemark{a}} &
  \colhead{$\Delta$ Dec.} &
  \colhead{$\sigma_{Dec}$\tablenotemark{a}} &
  \colhead{v$_{RA}$} &
  \colhead{$\sigma_{v_{RA}}$\tablenotemark{b}} &
  \colhead{v$_{Dec}$}  &
  \colhead{$\sigma_{v_{Dec}}$\tablenotemark{b}}   &
  \colhead{Standard\tablenotemark{c}} &
  \colhead{PSF Star\tablenotemark{d}} \\
   \colhead{} &
  \colhead{(mag)} &
  \colhead{(year)} &
  \colhead{(arcsec)} &
  \colhead{(arcsec)} &
  \colhead{(mas)} &
  \colhead{(arcsec)} &
  \colhead{(mas)} &
  \colhead{(mas yr$^{-1}$)} &
  \colhead{(mas yr$^{-1}$)} &
  \colhead{(mas yr$^{-1}$)} &
  \colhead{(mas yr$^{-1}$)} &
  \colhead{} &
  \colhead{}
}
\startdata
      S0-6  & 14.3  & 2010.86  & 0.36  &   0.0154  &   0.2000  &  -0.3563  &   0.2200  &   -5.057  &    0.051     &    3.065     &    0.075     &      Y     &                 \\
     S0-11  & 15.4  & 2011.37  & 0.49  &   0.4896  &   0.2200  &  -0.0657  &   0.4300  &   -3.637  &    0.033     &   -2.530     &    0.080     &      Y     &                 \\
      S0-7  & 15.4  & 2012.37  & 0.54  &   0.5268  &   0.2600  &   0.1002  &   0.2900  &    5.711  &    0.052     &    0.727     &    0.069     &      Y     &                 \\
     S0-13  & 13.5  & 2011.00  & 0.69  &   0.5575  &   0.2000  &  -0.4073  &   0.2400  &    1.874  &    0.048     &    3.136     &    0.063     &      Y     &                 \\
     S0-12  & 14.4  & 2011.53  & 0.69  &  -0.5497  &   0.2200  &   0.4198  &   0.2100  &    1.060  &    0.045     &    3.399     &    0.055     &      Y     &             C\_SW \\
     S0-31  & 15.1  & 2012.08  & 0.73  &   0.5810  &   0.4400  &   0.4470  &   0.8200  &    6.377  &    0.116     &    0.885     &    0.273     &      N     &             C\_SW \\
     S0-14  & 13.9  & 2011.40  & 0.81  &  -0.7533  &   0.2200  &  -0.2888  &   0.2800  &    2.514  &    0.051     &   -1.446     &    0.082     &      Y     &      C\_NW,C\_SW,W \\
      S1-5  & 12.7  & 2011.08  & 0.94  &   0.3142  &   0.2400  &  -0.8833  &   0.1800  &   -3.671  &    0.067     &    4.304     &    0.053     &      Y     &                 \\
\enddata
\tablecomments{Table 3 is published in its entirety in the machine-readable format.
      A portion is shown here for guidance regarding its form and content.}
\tablenotetext{a}{Positional errors include centroiding, alignment, and residual distortion(1 mas) errors, but do not include error in position of Sgr A* (0.01 mas ad -0.16 mas in RA and Dec respectively.}
\tablenotetext{b}{Velocity errors do not include error in velocity of Sgr A* (0.018 mas yr$^{-1}$ and 0.004 mas yr$^{-1}$ in RA and Dec respectively.}
\tablenotetext{c}{Indicates whether the star is a secondary astrometric standard.}
\tablenotetext{d}{Indicates for which mosaic field the star was used to create a PSF.}
\label{tab:secondary}
\end{deluxetable*}

\begin{figure*}[t!]
\epsscale{1.1}
%\vspace{-20pt}
\plotone{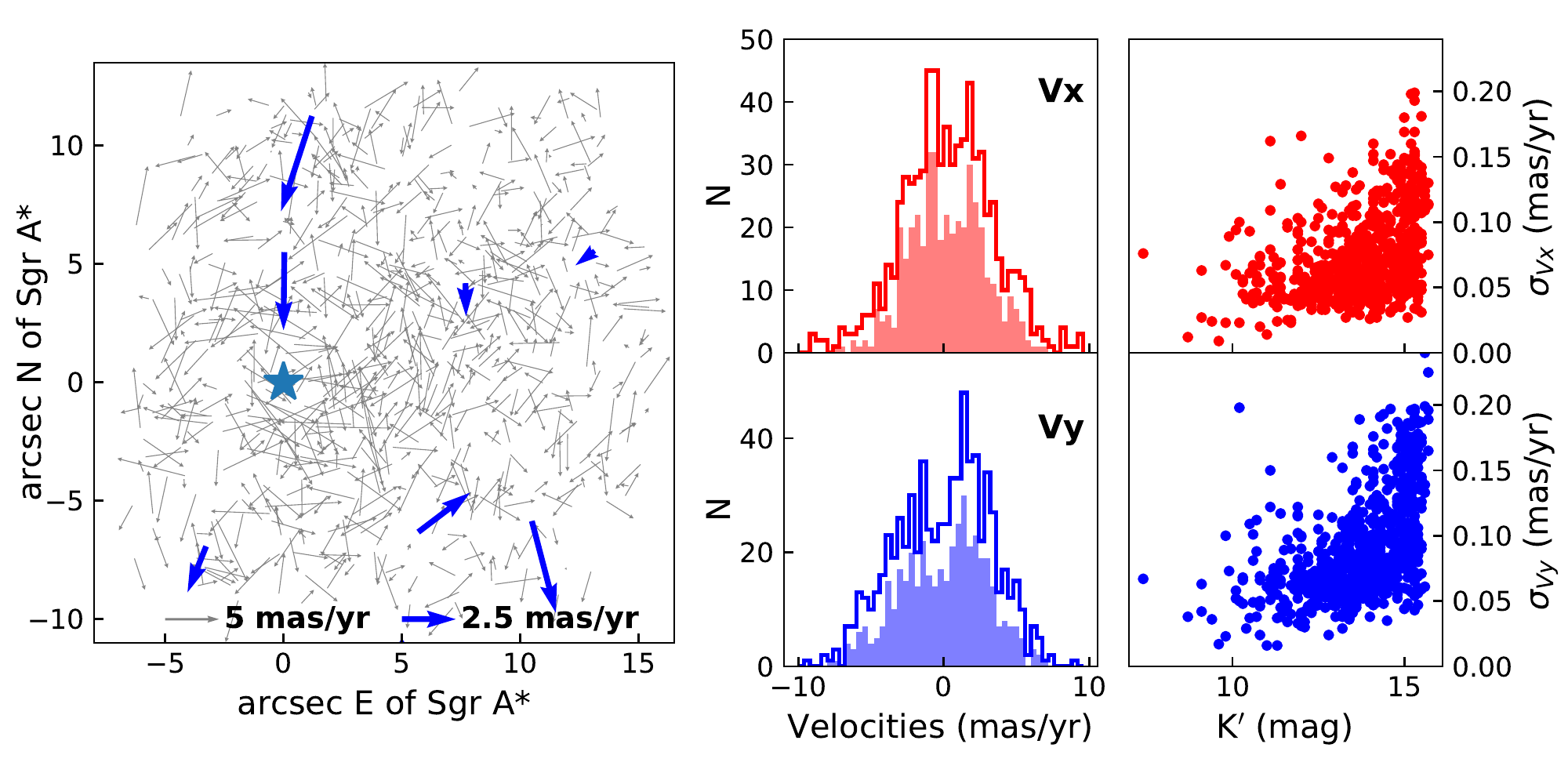}
\caption{
{\it Left: } The distributions of proper motions on the sky. Seven masers highlighted with blue arrows and the position of Sgr A* by a star are shown. There is no obvious pattern seen in the distribution; the direction of proper motion vectors appear to be directed randomly.  
{\it Middle:} The number distributions of proper motions.  Skewness is seen in the Y  direction which is likely due to rotation along the Galactic plane seen preferentially on the near-side of the Galactic Center. The unshaded histograms correspond to the distributions of all stars, while the shaded histograms show those of stars outside the 7$\arcsec$ radius.
{\it Right:} The distributions of proper motion uncertainties. 
The uncertainties in the Y direction are worse for a given magnitude.  This is likely due to the position of the tip-tilt star used in the NIRC2 observations.  
\label{fig:velocities_quiver}}
\end{figure*}

We have also examined the distribution of the directions of proper motion vectors as shown in the left panel of Figure~\ref{fig:velocities_quiver}.  The magnitudes and directions of all stars used as secondary astrometric standard stars are plotted here, with seven masers highlighted by thicker blue arrows.  There is no obvious pattern in the distribution of these vectors; they are randomly spread, suggesting that there is likely no underlying directional bias in the method used to derive proper motion measurements of stars in the GC region.

In the middle panel of Figure~\ref{fig:velocities_quiver}, the proper motion distributions in the East and North direction are shown in the top and bottom panel respectively.
Skewness in the velocity distribution is observed, especially in the N/S direction. This is likely due to the fact that the early-type stars that comprise the clockwise disk of stars are included in the sample.  Outside a 7$^{"}$ radius from Sgr A*, although the skewness in the velocity distribution decreases slightly, there are clearly more stars with positive velocities than negative velocities.
Furthermore, in the right panels of Figure~\ref{fig:velocities_quiver}, the distribution of proper motion errors is shown.  The velocity errors in the N/S direction are slightly larger than those in the E/W direction on average.  It was also already shown that the agreement between the IR and radio proper motions in the Y direction is not as well constrained as it is in the East direction, as seen from the average values listed in Table~\ref{tab:maser_polyfit}.  As mentioned above in Section~\ref{sec:starlists}, this is likely due to the fact that the tip tilt star used for the LGSAO observations is to the northeast position of the maser mosaic NIRC2 fields, resulting in the larger uncertainties in the geometric distortion correction in the North direction.  In the middle panel of Figure~\ref{fig:velocities_quiver}, the velocity distributions are shown.  The mean position and velocity of 286 stars within 7$\arcsec$ are zero within 1$\sigma$.  The velocities of the same set of stars within 7$\arcsec$ are also zero.

% on laptop ~/Galactic Center/RefFrame/AbsoluteRef2017/H_ver2_jknife/new_label_results/paper_plots.py

\begin{figure*}[t]
\plotone{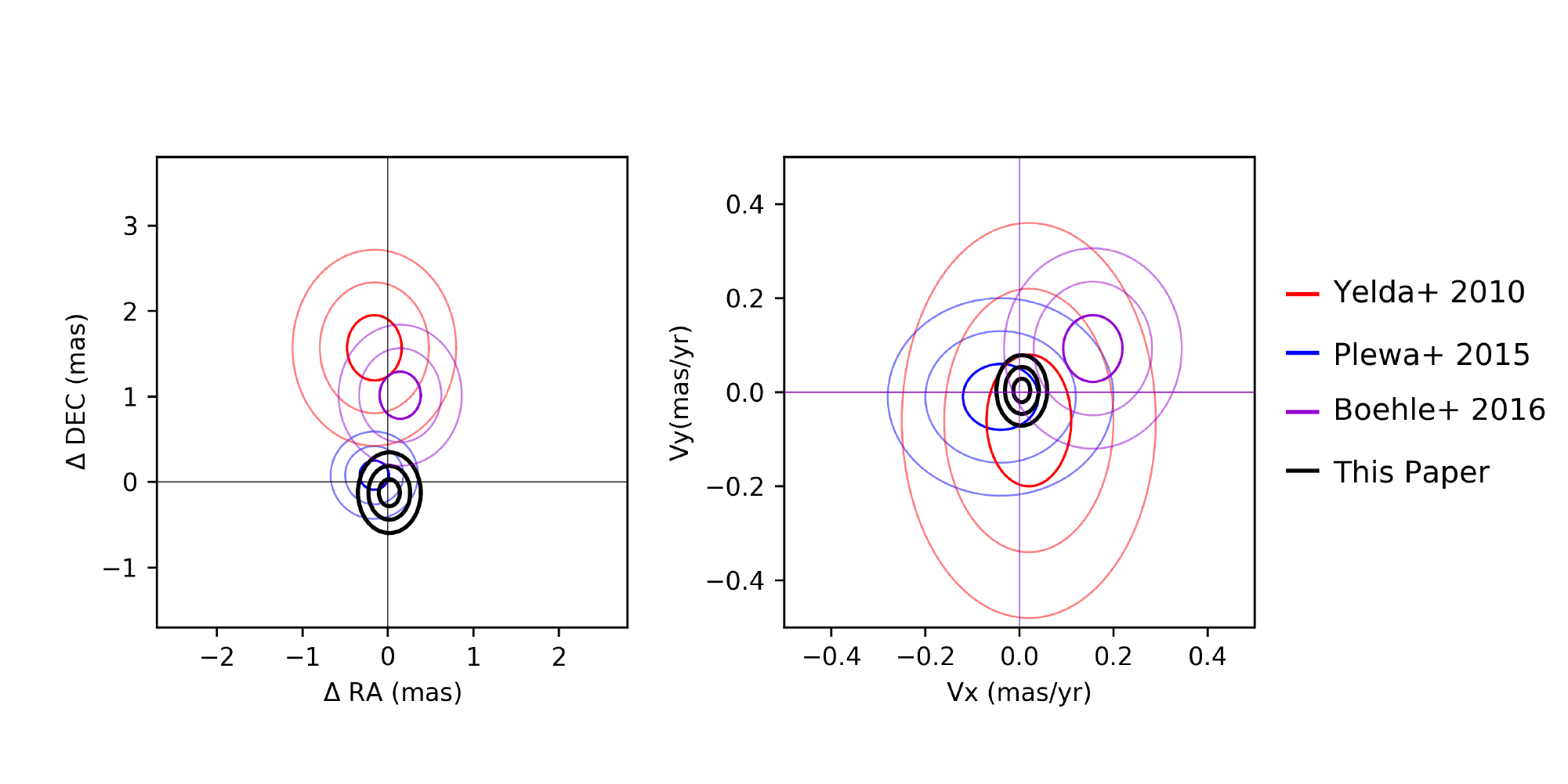}
\caption{Comparison of the stability of the reference frame as measured by the average difference between the positions and velocities in the radio and IR.  The agreements of the maser positions ({\it left}) and velocities ({\it right}) in the IR are compared to those in the radio.
The ellipses representing 1, 2 and 3 $\sigma$ uncertainties. 
For the Keck/NIRC2 data, the uncertainty in the distortion correction of 1 mas which dominates the IR error was subtracted as to show the improvement in the methods used in this paper to construct the reference frame.
%\textcolor{red}{[JLU: I think we need to state what metric of "stability" we are using here. Should it be "as measured by the average difference between the positions and velocities in the radio and IR"? ]}
\label{fig:coord_sys}}
%/u/shoko/code/python/refframe/coord_sys.py
\end{figure*}

\subsubsection{Comparison with Previous Works}

In Figure~\ref{fig:coord_sys}, the stability of the reference frame derived in this paper is compared with those of previously-published results.  
In \cite{Yelda:2010ig}, six epochs of maser data were used, instead of 13 used in this paper.  When comparing the stability of the reference frame, the distortion correction uncertainty of 1 mas that was added in the formal IR error (see Table~\ref{tab:maser_polyfit}) is subtracted, as we would like to focus on the improvement made in the methods used to construct the reference frame. 
Without this distortion uncertainty, the \cite{Yelda:2010ig} reference frame yields  uncertainties in Sgr A*'s position of 0.319 mas and 0.382 mas in the East and North directions, respectively.  In this paper, Sgr A*'s position is estimated with uncertainties of 0.122 mas and 0.157 mas in the East and North directions, respectively, indicating that the reference frame has been improved by a factor of $\sim$2.5 in position.
%Compared to \cite{Yelda:2010ig} in which 6 epochs of maser data were used instead of 13 used in this paper, the stability of the reference frame has improved by a factor of $\sim$5; 
The uncertainties in the proper motions of Sgr A* in the IR reference frame are 0.02 and 0.03 mas yr$^{-1}$ in the East and North direction respectively, compared to 0.09 and 0.14 mas yr$^{-1}$ reported by \cite{Yelda:2010ig}.  
The orbital analysis of \citet{Boehle:2016ko} was based on the reference frame which had three epochs of maser observations (2011-13) in addition to those used in \cite{Yelda:2010ig}. The uncertainties in the proper motions in the East and North direction of Sgr A* were 0.056 and 0.060 mas yr$^{-1}$ respectively.
%, corresponding to an improvement of a factor of $\sim$2.

Using the NACO imager on VLT, \cite{Plewa2015} reports a stability of $\sim 0.17$ mas in position and $\sim$0.07 mas yr$^{-1}$ in velocity.  Their analysis used eight masers as their mosaicked field covered 42 arcsec x 42 arcsec.  However, the  \cite{Plewa2015} maser sample did not include IRS7.  
IRS7 is a supergiant and its SiO maser features originate from a much larger maser emission regions than the other masers used \citep{Reid2003} by a factor of $\sim 2$, which is reflected in a larger uncertainties in the star's radio positions and proper motions.  To see how much effect this one star has on the IR reference frame, we construct the reference frame without IRS7.   Excluding IRS7, we obtain combined positional and velocity uncertainties of 0.17 mas and 0.031 mas yr$^{-1}$ respectively.   The velocity stability of our sample of six masers, without IRS7, is still a factor of 2 better than the one reported by \citet{Plewa2015} This is because IRS7 has less influence on the overall reference frame due to the very large errors assigned to its radio velocities.  
The maser sample used by \cite{Plewa2015} also had one fewer epoch of the radio observations. Furthermore, their IR observations included data through 2013, while our IR data extends four additional years.  As explained in \ref{sec:discussion}, the IR data and the method used to create the reference frame presented in this paper were modified significantly compared to those used in \cite{Yelda:2010ig}.  The comparison of the \cite{Boehle:2016ko} with that of \cite{Plewa2015} is likely more appropriate, as both data use the maser data up through 2013.  

\section{Discussion} \label{sec:discussion}

We have presented an improved astrometric reference frame that utilizes 13 years of IR observations of radio-emitting SiO masers in the vicinity of the Galactic Center. Since our previous work in \citet{Yelda:2010ig}, several modifications have been made in  the reference frame construction including: 
using 13 epochs of maser data (2005--2017) instead of 6 epochs (2005--2010) used by \cite{Yelda:2010ig};
using an improved PSF-fitting package, AIROPA, instead of StarFinder v1.6 \citep{Diolaiti:2000tz} used in \citet{Yelda:2010ig};
usinga new method to mosaic 9 or 13 dither positions for each epoch of observations.  \citet{Yelda:2010ig} stitched together one field at a time, building up the starlist by one position after another.  Instead, we transformed all dither positions simultaneously by transposing each list to the reference master starlist;
applying a six-parameter fit to transform the IR maser positions to the radio Sgr A*-rest reference frame, while \citep{Yelda:2010ig} used a four-parameter fit; and 
using an updated set of radio maser positions and proper motions (Reid, private comm).

%FROM MARK REID -- NEED TO PUT THIS IN THE FIGURE pos_vel.eps
%In that paper we give the RAcos(Dec) motion as -3.151 +/- 0.018 mas/yr and the Dec motion as -5.547 +/- 0.026 mas/yr.
%I have unpublished updated values of -3.156 +/- 0.006 and -5.585 +/- 0.010 mas/yr, based on a much longer time baseline.
%{\it  the bottom line is that \#1 and \#5 have the most effect.  Other factors do not seem to change the ref frame that much- not significantly at all.  \#3 does as we could not really do the cross-epoch alignment before adapting the higher order polynomial to make each master starlist.  Table 6 shows some of the tests I've done but this was done by going 'backward' - starting with the CURRENT ref frame and changing one thing - like starfinder instead of AIROPA but everything else kept the same - not starting with Yelda et al and changing one thing and see how much it improved.  So not sure if this is telling us that much.  But if each one has a few \% effect, then the combined effect would be quite large?}

\subsection{Dependence on the Number of Epochs of Observations}

The stability of the reference frame improved by a factor of $\sim$5 compared to that of \citep{Yelda:2010ig}, to which the change in each one of the above processes contributed.   Of the five, the number of epochs of maser mosaic observations has had the most significant effect on the improvement of the reference frame stability.   If we were to build the reference frame using seven epochs of data spanning the years 2005--2011, with all other conditions remaining the same, the uncertainties in the reference frame velocities would increase to 0.023  mas yr $^{-1}$ and 0.043 mas yr$^{-1}$ for $V_{RA}$ and $V_{DEC}$ (from 0.018 and 0.025 mas yr$^{-1}$).   However, if we used seven epochs spread over 13 years from 2005 through 2017, skipping every other year, then the velocity uncertainties would remain similar (0.021 mas yr$^{-1}$ and 0.025 mas  yr$^{-1}$).  Thus it is not just the number of epochs that is important in creating a stable reference frame, but the range of dates of observations.  The longer the time baseline, the more accurate the reference frame becomes.  We attribute this to the improved precision and accuracy of the proper motions of the secondary astrometric stars.

\begin{figure*}[hbt!]
%on laptop  %~GalacticCenter/cAbsoluteRef2017/H_ver2_jknife/DOM/Comparison/contour_all.py
\epsscale{0.8}
\plotone{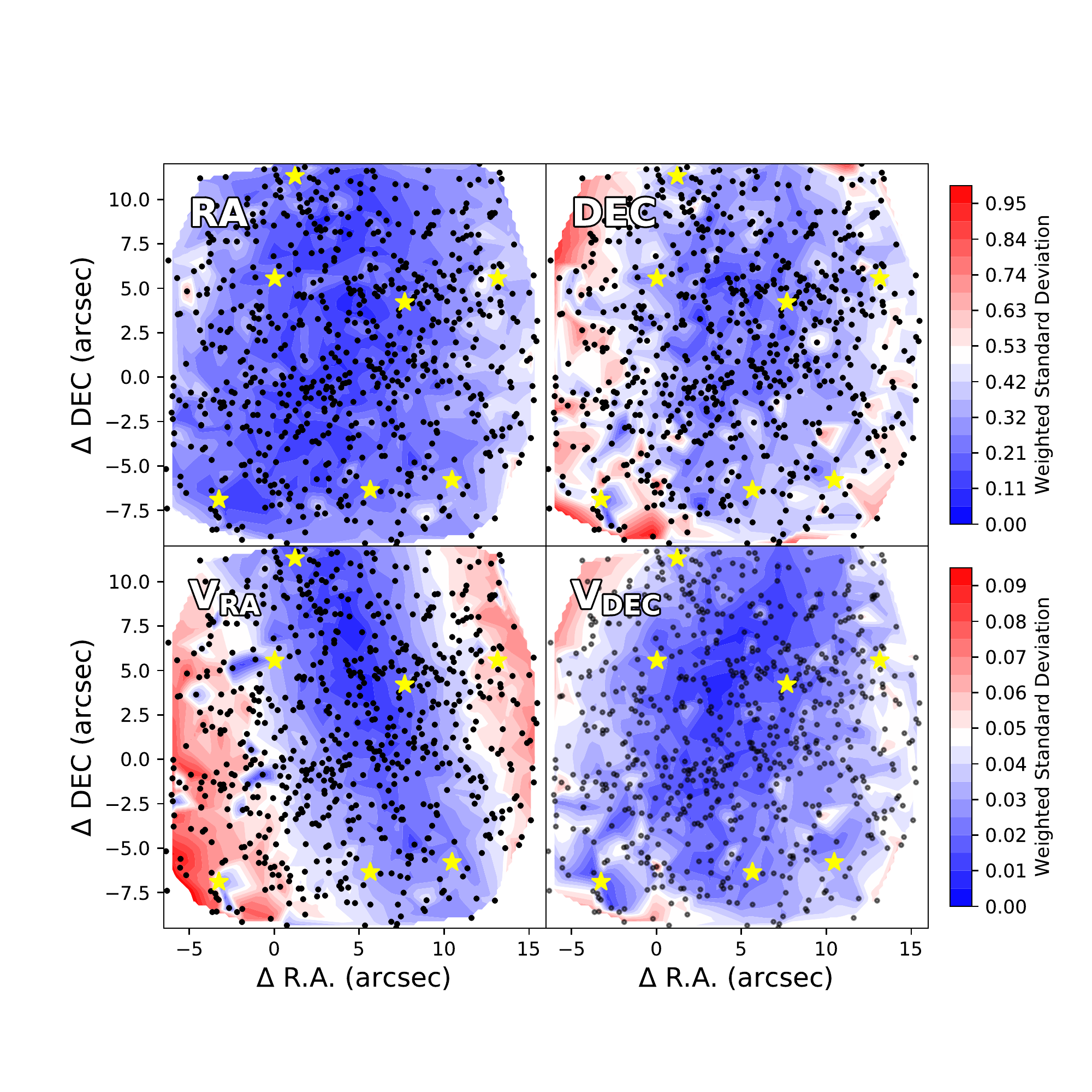}
\caption{Stability of the stellar positions ({\em top}) and velocities ({\em bottom}) of the secondary astrometric standard stars as determined from a jackknife bootstrap over the seven masers.  The dots show the positions of secondary astrometric stars.  The positions of masers are shown by yellow stars.  For each star, shifts in the R.A. and DEC positions for the year 2000.0 and $V_{RA}$ and $V_{DEC}$ proper motions as you drop one maser from the reference frame construction are calculated and the weighted  standard deviations of seven drop-one-maser cases are calculated and plotted. The shaded color map shows the weighted standard deviation values in RA (upper left) and DEC (upper right) positions and RA (lower left) and DEC (lower right) proper motions. 
%\textcolor{red}{Add label to color bar and fix X-axis for flipping?. Make bigger dots for the masers.} 
\label{fig:contours_xy}}
\end{figure*}

\subsection{Stability of the Reference Frame Based on the Choice of Masers}
% DROP ONE MASER TEST
The IR astrometric reference frame is stable within 0.03 mas yr$^{-1}$ (Table~\ref{tab:maser_polyfit}).  However, this is based on the radio observations of proper motions of seven SiO masers only. 
There are 16 masers that can be used potentially to create the Sgr A*-radio rest reference frame \citet{Reid2003}.  However, given the field of view of NIRC2 and the telescope scheduling feasibility, we have only been able to mosaic together the field large enough to cover seven masers closest to Sgr A*.
Because of its dependence on a low number of astrometric anchor points, we examine in this section how sensitive the global parameters of the SMBH are to the selection of masers. 

We have applied a jackknife resampling method, in which one maser at a time is excluded from the construction of the reference frame.
The proper motions of secondary astrometric standard stars derived by the radio positions of the remaining six masers change systematically. As mentioned above, the positional uncertainties of secondary astrometric stars are calculated by taking the average of seven drop-one-maser cases, with the assumption that this does not lead to any systematic uncertainties that might depend on the location of the star on the maser mosaic field.  In order to evaluate whether this is the case, we have estimated first the deviation of the astrometric and proper motion values of one of the jackknife cases compared to those in the case in which all seven masers were used.  The weighted standard deviation of seven cases were then calculated and plotted by color maps in Figures~\ref{fig:contours_xy}.  For the astrometric comparison in Figure~\ref{fig:contours_xy}, the positions in the epoch 2000.0 were used.  

\begin{figure*}[htb!]
\epsscale{1.1}
\plotone{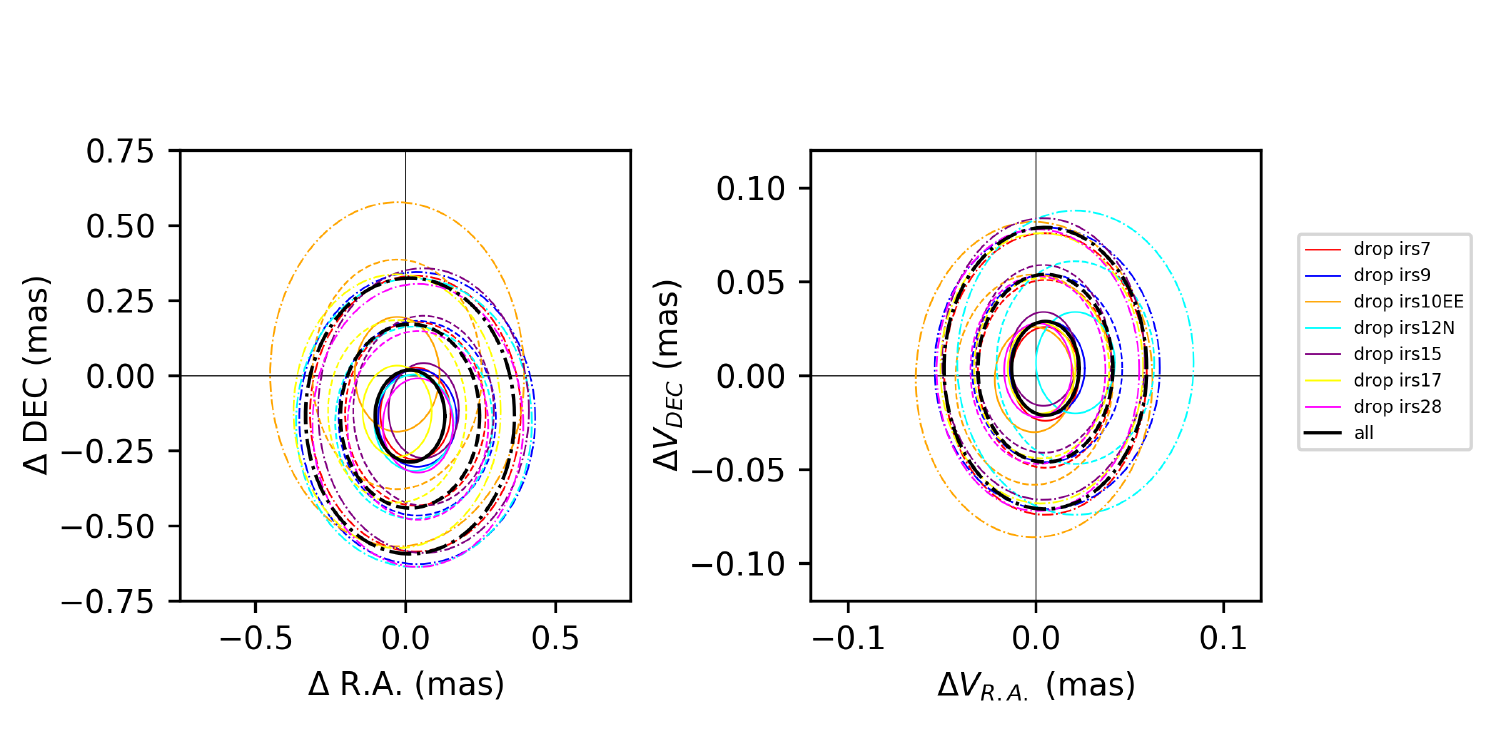}
\caption{Comparison of the IR reference frame determined for each drop one maser case with the radio astrometric reference frame.     
For each maser dropped, 1, 2 and 3-$\sigma$ uncertainties are shown.
The dropped maser is color-coded following the legend on the right.
The radio astrometric coordinate system is based on the assumption that the Sgr A* is at (RA,DEC)=(0,0) and at rest.  \label{fig:dom_xy_vxvy}}
%/u/shoko/code/python/refframe/coord_comp.py
\end{figure*}

%The stability in the parameters of the SMBH is also examined.  
In Figure~\ref{fig:dom_xy_vxvy}, the comparisons of IR and radio positions of masers in all seven jackknife cases are shown.  The case in which all masers are used is also included in this figure as the "all" case.  
Each jackknife case agrees with each other, and also with the "include all" case within 1$\sigma$, suggesting that the selection of masers should not affect the zero point of the IR astrometric reference frame.   
We further investigate how the SMBH parameters may be affected by the maser selection in the next section.

\subsection{The Effect of the New Reference Frame on the S0-2 Orbit}

\begin{figure*}[t!]
\epsscale{1.1}
\plotone{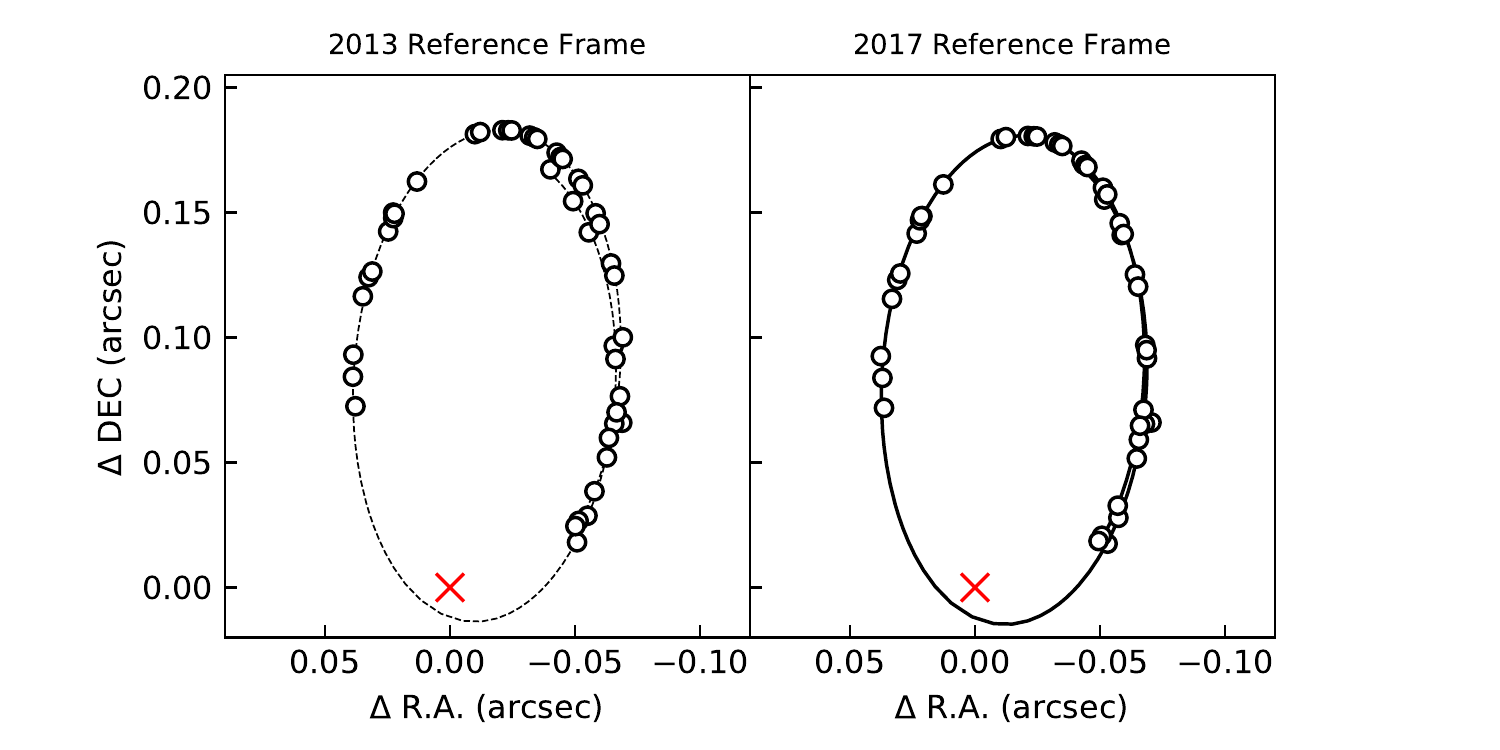}
\caption{
%\textcolor{red}{JLU: Mark the position of the black hole. Stop the orbit at the last data point (don't %extrapolate). } 
Comparison of S0-2 orbits based on the current IR reference frame presented in this paper ({\em right}) with the one presented in \citet{Boehle:2016ko} ({\em left}).  The position of the black hole is indicated by a red cross.
\label{fig:orbits_comp}}
\end{figure*}
%/local/code/python/refframe/Papers/orbit_comp.py

Since the IR positions of all stars on the master starlist per epoch are transposed to the radio Sgr A*-rest frame, if the perfect  reference frame is achieved, the mean position and proper motion of Sgr A* in the IR reference frame should be zero.  If not, the orbit of for e.g., S0-2, would exhibit a shift resulting in an unclosed orbit, as the zero point of each coordinate system would systematically wander.  
For the reference frame being presented in this paper, as seen in Table~\ref{tab:maser_polyfit}, the agreement between the IR and radio maser positions and proper motions is consistent with zero within uncertainties.  
In Figure~\ref{fig:orbits_comp}, the positions of the 16-year-period star S0-2, spanning the period of 1995 - 2017, after being cross-epoch aligned to the common reference frame are shown (Jia et al 2018) in the right panel, with the orbit model fit superposed in black.  
Plotted in the left panel in the same figure is the S0-2 astrometric points cross-epoch aligned using the reference frame used in \citet{Boehle:2016ko}, with the orbit model fit superposed.  The same set of NIRC2 and radial velocity data were used in both orbits shown in Figure~\ref{fig:orbits_comp}; the only difference between the two orbits is the reference frame used.
With the previous reference frame, the shift in the radial position of S0-2 was roughly 0.75 mas yr$^{-1}$which is clearly seen in Figure~\ref{fig:orbits_comp} as the gap in the orbit in the northeast direction from Sgr A*, whereas with the newer current reference frame, the same shift is $\sim$0.05 mas yr$^{-1}$, an order of magnitude better. Since the \citet{Boehle:2016ko} analysis, the speckle holography data have been re-analyzed and been updated.  The orbit-fitting results shown in this paper utilize the updated newer version (Jia et al 2018).  However, when estimating the orbit of S0-2 using the previous version of speckle holography data, but using the most current reference frame being presented here, the orbit is much more "closed", similar to the one shown in Figure~\ref{fig:orbits_comp}, compared to the one presented by \citet{Boehle:2016ko}, suggesting that the new reference frame plays a larger role in refining the S0-2 orbit compared to the re-analyzed NIRC2 speckle data. 

We further examine how each jackknife reference frame affects the black hole parameters determined from the orbit fitting following the procedure described in detail in \citet{Ghez:2005ck,Ghez:2008ty,Boehle:2016ko}. 
%Seven independent absolute IR reference frames were created, each corresponding to the case in which one maser is dropped from the analysis.  
Each jackknife reference frame, as described in the previous Section, is used for cross-epoch alignment of Central 10 data, followed by orbit-fitting procedure for each case to estimate the parameters of the Galactic Center SMBH \citep[see][]{Ghez:2008ty,Boehle:2016ko}. 
The case in which no maser is dropped is the same alignment of cross-epoch Central 10\arcsec\; data presented in Jia et al (in preparation).  
Examining the statistical bias of SMBH parameters using the seven subset reference frames by applying the equations in the Appendix C of \cite{Boehle:2016ko}, we estimate that the measurements are biased at a 2$\sigma$ level.

\begin{figure}[ht!]
\epsscale{1.2}
\plotone{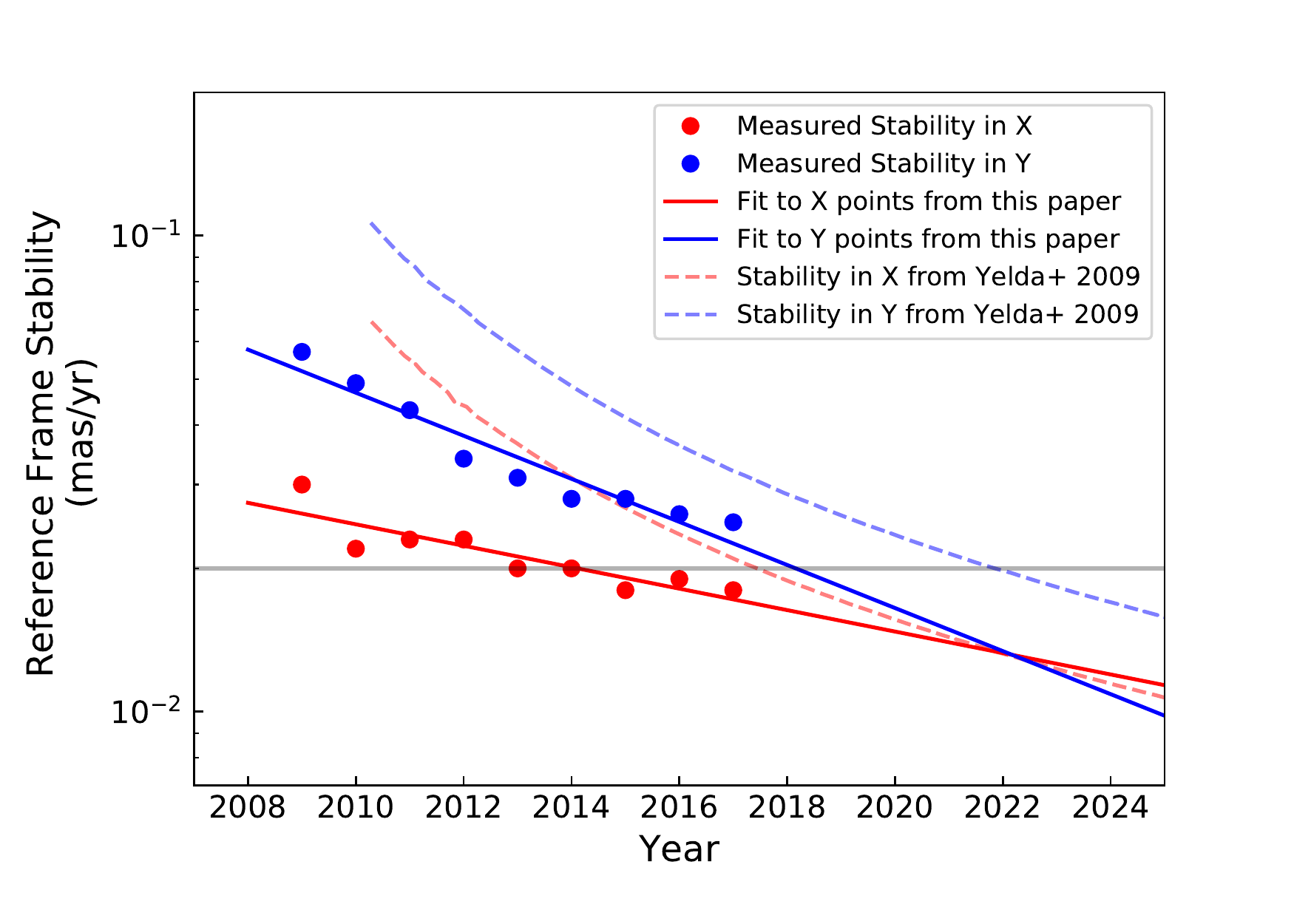}
%on workstation:  ~/AbsoluteRef2017/H_ver2_jknife_EPOCHS/
\caption{
%\textcolor{red}{JLU: Label the lines in the legend.} 
The improvement in the stability of reference frame as a function of time, which corresponds to the number of epochs of observations.  We have obtained one set of maser observations every year starting in 2005. As of 2017, 13 epochs of observations in IR have been made. The figure shows the reference frame stability determined for each epoch, using maser observations up to and including that epoch.  For future points, it is assumed that one set of maser observations is taken every year. The actual observed X and Y proper motion errors are represented by red and blue solid circles respectively. The solid lines show the fits through the observed data points, while the dashed lines show the predictions from \citet{Yelda:2010ig}. 
%\textcolor{red}{The fonts seems a little too small on the labels for this figure.
The gray horizontal line shows the uncertainty that needs to be achieved in order to observe the apocenter shift of S0-2.
\label{fig:errors}}
\end{figure}

One of the motives for improving the IR reference frame has been the opportunity to be able to test General Relativity, as S0-2 reached its closest approach to the SMBH in 2018.  Furthermore, in the future, the observations of the apocenter shift of S0-2 should be possible, if the uncertainty in the reference frame stability of $\sim$0.02 mas yr$^{-1}$ is achieved (\citet{Weinberg2005}).
\citet{Yelda:2010ig} reported that this stability would not be achieved until $\sim$2022 based on six epochs of maser observations.   We are now able to revisit this question, with the data that is more than double in size.  Figure~\ref{fig:errors} shows the improvement in the reference frame.  It displays the uncertainty in the Sgr A* velocity in the IR reference frame as a function of time.  The trends presented in \citet{Yelda:2010ig} are represented in this Figure by dashed lines.  By year 2020, with three additional epochs of maser data starting in 2018, the combined stability of $\sim 0.02$ mas yr$^{-1}$, shown by thin gray horizontal line, can be reached.

\subsection{Position of Sgr A*}

\begin{figure}[ht!]
\epsscale{1.2}
\plotone{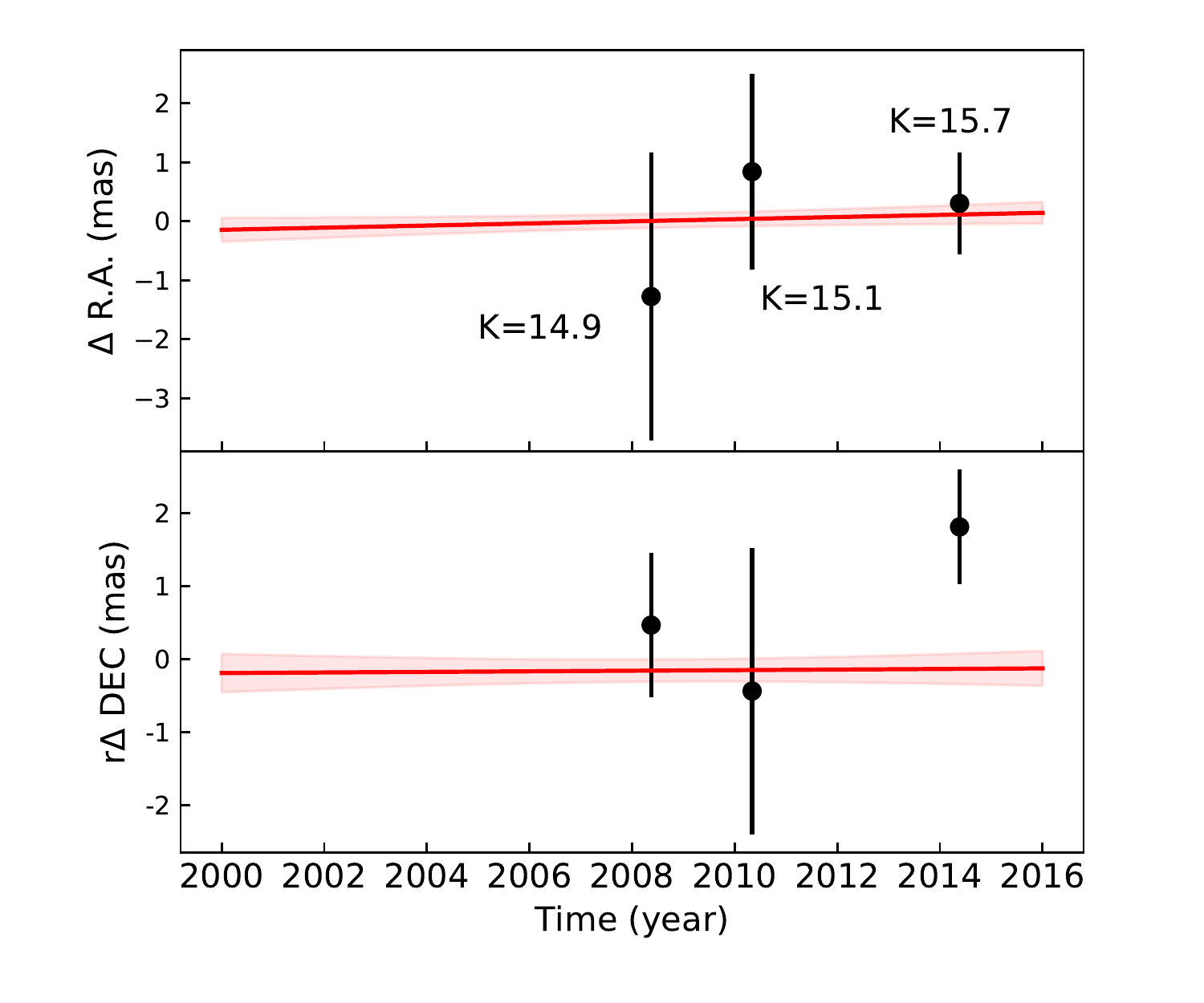}
\caption{R.A. and DEC positions of Sgr A*-IR as detected in the maser mosaic image.  The {\em red line} shows the predicted positions and uncertainties determined from the reference frame as shown in Table~\ref{tab:maser_polyfit}. \label{fig:sgr_pos}}
%/u/shoko/Data2017/For_Sgr/
\end{figure}

Another method of examining how well the IR astrometric reference frame is determined is to compare the position of Sgr A*-IR on maser mosaic frames as compared to the predicted Sgr A* radio position.  
Genzel 2003, Ghez 2004, 2005
Sgr A*-IR is highly variable (\cite{Genzel2003}, \cite{ghez2004variable}, \cite{ghez2005redemission}, \cite{Witzel:2017vk}), and it is not often detectable, especially in short-exposure maser mosaics.  
However, we make use of the Central 10'' data by examining single frames from a given epoch, which allows us to pinpoint the location of Sgr A*-IR when it flares.  
The position of Sgr A*-IR is then located on the maser mosaic field if the maser and Central 10$^{"}$ observations were taken within approximately a month of each other.  Otherwise, the stars closest to the SMBH would have moved enough such that the location of Sgr A*-IR cannot be determined accurately enough.
We have examined each epoch of data and found that Sgr A*-IR is visible on three epochs: 2008, 2010 and 2014.  The results are shown in Figure~\ref{fig:sgr_pos}.   The RA and DEC positions of Sgr A*-IR in three epochs are shown as a function of time, with the predicted positions and uncertainties determined from the reference frame, as shown in Table~\ref{tab:maser_polyfit}, over-plotted.  With the exception of the 2014 DEC position, the Sgr A*-IR position agrees well with the zero point of the reference frame.  The brightness of Sgr A*-IR is fainter in 2014 than in other two epochs, which may explain the slight discrepancy.  

%Consider adding the following somewhere in the paper: 
%\textcolor{red}{How will Gaia help the reference frame? Could we include this here? -- yes}  
%\textcolor{red}{Can we make a table showing the astrometric error budget for future periapse precession experiments and show how the reference frame factors in?}

\section{Summary}

We have presented an improved astrometric reference frame in the near-IR, which is used for monitoring the positions of stars in the vicinity of Sgr A* to determine the SMBH properties.  In the current IR reference frame, the position of Sgr A* is localized within 0.113 mas and 0.147 mas in the East and North positions respectively, and 0.008 mas yr$^{-1}$ and 0.004 mas yr$^{-1}$ in the East and North proper motions respectively.

As shown in \S\ref{sec:discussion}, the  IR astrometric reference frame still depends slightly on the choice of the masers; for example, the IR position of  Sgr A* is estimated with bias of 1.9 - 2.3 mas.   The solution to decrease this bias may be to include additional masers in the analysis. However, the observation of additional NIRC2 positions is time-consuming, and realistically not viable. There are several GAIA sources with proper motion measurements in the Galactic Center region. However, again, they are unfortunately located just outside our maser mosaic field. Mutli-epoch observations with Hubble Space Telescope should be able to provide additional stellar positions and proper motion data within the mosaic field. Any stars can be used as reference stars, as long as their proper motions are estimated accurately enough.

\section{Acknowledgements}

We thank the staff of the Keck Observatory, especially Randy Campbell, Jason Chin, Scott Dahm, Heather Hershey, Carolyn Jordan, Marc Kassis, Jim Lyke, Gary Puniwai, Julie Renaud-Kim, Luca Rizzi, Terry Stickel, Hien Tran, Peter Wizinowich, Carlos Alvarez, Greg Doppman and current and former directors, Hilton Lewis and Taft Armandroff for all their help in obtaining observations.
We would also like to thank the anonymous referee for very useful comments and suggestions. 
Support for this work at UCLA was provided by 
Heising Simons Foundation (2017-282), and
National Science Foundation (AST-1412615).
Also, matching funds to the The W.M. Keck Foundation (20170668) were provided to UCLA and UC Berkeley. Furthermore, S. J.,  J.R.L., and M. W. H. acknowledge support from NSF AAG (AST-1518273).

The W.M. Keck Observatory is operated as a scientific partnership among the California Institute of Technology, the University of California and the National Aeronautics and Space Administration.  The Observatory was made possible by the generous financial support of the W. M. Keck Foundation.  The authors wish to recognize and acknowledge the very significant cultural role and reverence that the summit of Maunakea has always had within the indigenous Hawaiian community.  We are most fortunate to have the opportunity to conduct observations from this mountain.

\facilities{Keck Observatory}

\software{
astropy \citep{astropy},
KS2 \citep{Anderson2008},
Matplotlib \citep{Hunter:2007},
SciPy \citep{scipy}
}

\appendix
\section{Local Distortion Correction}
\label{app:ldc}
In \citet{Yelda:2010ig}, four-parameter fits (X/Y translation, scale, and rotation) were used to mosaic together stellar positions of nine dither images  to create one starlist for each epoch. 
%XXX WHAT IS THE ORDERING OF THESE PAIR-WISE FITS, that could matter ?xxx
However, we have adopted a third-order polynomial fit in Step (1) in Section 2.4 when matching the observed astrometric positions to the master starlist, to accommodate the change in the optical distortion corrections in 2015 \citep{Service2016}, caused by the re-alignment in the AO system and NIRC2 camera. Before the 2015 change, there was no evidence of time variability in the distortion solution, which captures the geometric optical distortions and likely some PSF variation over the field.  %Not needing further corrections that are time-dependent suggested that everything prior to 2015 (e.g. geometric and phase aberrations).  However 
%After the 2015 AO alignment change, the need to apply a higher-order, time-dependent local distortion correction to stellar positions on each mosaic dither position emerged (Jia et al. 2018), likely due to the work that has been done continuously on the AO system at Keck Observatory. 
\cite{Service2016} only had one year of time sampling and the distortion seemed constant within the uncertainties. However, we have subsequently found that there are time-dependent variations in the distortions beyond 2015.  This is apparent in the Central 10" pointings of the Galactic Center data, which are always taken at the exact same sky options.  We do not have enough data to determine if the distortion pattern is drifting slowly or is more randomly changing.  Thus we are unable to construct a "master" distortion map for all epochs.
%XXX or could it be that your measurements had just suddenly gotten more accurate to see subtle distortions previously missed ? XXX

The details of deriving the local distortion correction for the Central 10$^{"}$ field are presented in Jia et al (2018).  Briefly, it is determined from the residuals of the comparison between the observed position for that epoch and the predicted position from the proper motions from the standard NIRC2-LGSAO setup (Jia et al 2018).  A set of stars with high-accuracy proper motion measurements is required to create the local distortion map.   
Unfortunately, the local distortion correction for the central 10$^{"}$ field cannot be applied to the maser mosaic fields, as the correction appears to depend on the position of NIRC2 with respect to the tip-tilt star. The same tip-tilt star is used for the central 10$^{"}$ and maser mosaic observations. One of the major reasons for needing the local distortion correction is the PSF variation across the NIRC2 field.  Since the maser mosaic dither positions are all at different locations with respect to the Central 10$^{"}$ field, its PSF variation across the NIRC2 field differs as well.  Thus we cannot apply the Central 10$^{"}$ local distortion correction derived by Jia et al 2018 on maser mosaic stellar positions.

Furthermore, we are unfortunately not able to derive the local distortion correction for maser mosaic fields, since even though more than 700 secondary astrometric stars are found in the entire maser mosaic field, each dither position encompasses less than $\sim$100 such stars. This is not enough to fit a 4th order polynomial which would be needed to fully specify the local distortion correction.
Our solution is to compromise with a third-order polynomial fit to combine multiple maser fields together to create a master starlist for each epoch. 
%Given the local distortion correction requires a high-order polynomial fits of fourth or fifth order, a third-order fit applied to make the starlist seems reasonable.  
The residuals after the third-order fit are less than $\sim$0.1 mas, which is less than the error in the radio maser positions; thus we did not need to go to a higher order than we are currently using.
If we used the second-order parameter fit, as was done in \cite{Yelda:2010ig}, the post-2015 IR positions would appear significantly further away from the radio positions.
The median values of the geometric distortion correction are around 0.01 and -0.03 pixels in X and Y respectively before 2015 (\cite{Yelda:2010ig}) and -0.09 and 0.05 after 2015 (\cite{Service2016}), while the median values of the local distortion corrections are in the range of 0.02 - 0.05 pixels (~0.2 - 0.5 mas) with uncertainties around 0.01 - 0.04 pixels (~0.1 - 0.4 mas). 
These are relatively smaller than other sources of uncertainty discussed above.

\bibliography{refpaper}

\end{CJK*}

\end{document}